\documentclass[reprint, amsmath, amssymb, aps, showkeys]{revtex4-2}
\usepackage{graphicx}
\usepackage{dcolumn}
\usepackage{bm}
\usepackage{amsmath}
\usepackage{float}
\usepackage{subfigure}
\usepackage{booktabs}
\usepackage{CJKutf8}
\usepackage{hyperref}
\usepackage{tikz,xcolor}

\hypersetup{
  colorlinks=true,
  linkcolor=blue,
  urlcolor=blue,
  citecolor=blue,
  linktoc=all
}
\definecolor{lime}{HTML}{A6CE39}
\DeclareRobustCommand{\orcidicon}{%
	\begin{tikzpicture}
	\draw[lime, fill=lime] (0,0) 
	circle [radius=0.16] 
	node[white] {{\fontfamily{qag}\selectfont \tiny ID}};
	\draw[white, fill=white] (-0.0625,0.095) 
	circle [radius=0.007];
	\end{tikzpicture}
	\hspace{-2mm}
}

\foreach \x in {A, ..., Z}{%
	\expandafter\xdef\csname orcid\x\endcsname{\noexpand\href{https://orcid.org/\csname orcidauthor\x\endcsname}{\noexpand\orcidicon}}
}

\begin{document}
\begin{CJK*}{UTF8}{gbsn}

\title{Constraints and detection capabilities of GW polarizations\\with space-based detectors in different TDI combinations}

\author{Jie Wu (吴洁)\orcidA{} }
\author{Mengfei Sun (孙孟飞)\orcidC{} }
\author{Jin Li (李瑾)\orcidB{} }
\email{cqujinli1983@cqu.edu.cn}

\affiliation{College of Physics, Chongqing University, Chongqing 401331, China}
\affiliation{Department of Physics and Chongqing Key Laboratory for Strongly Coupled Physics, Chongqing University, Chongqing 401331, China}

\begin{abstract}
Time-delay interferometry (TDI) is essential in space-based gravitational wave (GW) detectors, effectively reducing laser noise and improving detection precision. 
Space-based GW detectors provide a unique opportunity to probe GW polarizations, including possible additional modes that may signal deviations from general relativity and alternative gravity theories. 
In this study, we examine the impacts of second-generation TDI combinations on GW polarization detection by simulating LISA, Taiji, and TianQin, including realistic orbital effects such as link length and angle variations. 
Detector performance is assessed using sensitivity and power-law integrated curves, as well as the signal-to-noise ratio (SNR) of binary black holes (BBHs) and phase transitions (PTs).
For massive BBHs, the $\mathcal{A}$ and $\mathcal{E}$ channels typically offer the best sensitivity, while the $X$ channel in TianQin is most effective for detecting additional polarizations. 
For stellar-mass BBHs, the $\alpha$ channel provides the highest SNR for vector modes in LISA and Taiji specifically for lower-mass systems, while the $\mathcal{A}$ and $\mathcal{E}$ channels are optimal for higher masses or other polarizations. 
For PT signals, the $X$ channel generally delivers the optimal performance, except in the low–peak-frequency regime below 1 mHz, where the $U$ channel in TianQin becomes more sensitive.
When considering additional polarizations, the $X$ channel emerges as the most robust choice for TianQin, in contrast to LISA and Taiji, where the $\mathcal{A}$ and $\mathcal{E}$ channels provide strong capabilities for GW polarization tests.
This distinction between LISA, Taiji, and TianQin represents a key result of the present work and has not been explicitly emphasized in previous studies.
Our findings emphasize the importance of selecting high-sensitivity TDI combinations to enhance detection capabilities across different polarizations, deepening our insight into GW sources and the fundamental nature of spacetime.
\end{abstract}

\maketitle
\end{CJK*}

% =======================================
\section{Introduction}
The first direct detection of gravitational waves (GWs) was made by LIGO and Virgo Collaborations in 2015, opening a new window for testing general relativity (GR) and exploring gravity in strong fields~\cite{GW150914}.
One way to test GR and constrain alternative gravity theories is to search for deviations from GR in GW observations~\cite{GR_deviation}.
Currently, more than 100 GW events have been detected, with no evidence of new physics beyond GR~\cite{test_GR1,test_GR2,test_GR3}.
Ground-based detectors like LIGO, Virgo, and KAGRA can only detect in frequency bands higher than 10 Hz due to their arm length limitations~\cite{aLIGO,aVIRGO,KAGRA}.
There are numerous sources in the low-frequency band, which facilitates more rigorous testing of GR~\cite{review_of_space_detectors}.
Constructing a space-based detector with an arm length of millions of kilometers in space can effectively detect low-frequency sources~\cite{my_paper2}.  
The proposed space-based GW detection missions, such as LISA~\cite{LISA}, Taiji~\cite{Taiji}, and TianQin~\cite{TianQin}, are designed to operate within a sensitive frequency band of millihertz, while DECIGO~\cite{DECIGO}, ALIA~\cite{Beyond_LISA}, and BBO~\cite{BBO_NS} are intended to cover a sensitive frequency band ranging from 0.1 to 10 Hz.

GW detection utilizes laser interferometry technology by precisely measuring the phase changes in the laser's round-trip propagation to capture GW signals.
Ground-based detectors can eliminate the laser frequency noise by establishing precise interferometric arms.
In contrast, the arms of space-based detectors change with time because orbital dynamics leads to variations in the distance between spacecraft (S/C).
Consequently, the laser frequency noise is several orders of magnitude higher than other noises, and it must be removed from the data to achieve GW sensitivity~\cite{TDI_LRR}.
To suppress laser frequency noise, Tinto~\textit{et al.} introduced time-delay interferometry (TDI), which constructs a virtual equal-arm interferometer by combining delayed scientific data streams~\cite{TDI_first}.
The combinations of TDI can be solved mathematically based on the relevant theories of algebraic geometry~\cite{TDI_theory}.
In reality, there might be a situation where one of the arms is damaged and data is missing, thereby resulting in the research and development of other combinations~\cite{TDI_different}. 

Some recent studies have discussed the TDI combination and GW polarization.
Wu~\textit{et al.} generalized the combinatorial algebraic approach to construct various classes of modified second-generation TDI solutions~\cite{TDI_all}.
In Ref.~\cite{TDI_theory}, Zhang~\textit{et al.} provided an analytical formula for the average polarization response function in different TDI combinations and obtained their asymptotic behaviors.
Wang~\textit{et al.} analytically evaluated the response functions for arbitrary TDI combinations while enumerating all possible polarizations~\cite{TDI_theory2}.
In Ref.~\cite{TDI_numerical}, Wang~\textit{et al.} used numerical methods to evaluate the noise level and sensitivity of different TDI channels.
These studies have provided valuable research directions and foundational insights for exploring GW polarizations.
Most of these studies focus either on theoretically deriving analytical or semi-analytical responses and sensitivities across different polarizations, or on numerically simulating tensor modes in various TDI combinations. 
Few have considered the scenario of simulating various space-based detectors probing different polarizations across diverse TDI combinations.

We conduct a detailed numerical study of the noise properties and polarization dependent responses of different TDI combinations by separately simulating LISA, Taiji, and TianQin. 
Building on our previous work~\cite{my_paper4,my_paper3}, which employed simplified detector responses to study GW polarizations, the present work extends the analysis to multiple TDI combinations and focuses on configurations that are more directly relevant to realistic space-based detectors. 
We perform a investigation of massive black hole binaries (MBHBs) and stellar-mass binary black holes (SBBHs) using different TDI combinations, and identify the optimal choices for different mass ranges. 
To characterize the detector performance in a broader context, we employ a stochastic gravitational wave background (SGWB) with a flat spectrum to model the GW response over the full frequency band and derive the corresponding sensitivity curves for different TDI combinations.
Based on these sensitivity curves, we further compute the power-law sensitivity for SGWB searches and evaluate the detectability of first-order phase transition (PT) signals under different TDI configurations.
Through this systematic investigation, we assess the constraints on GW polarizations and the detection performance from multiple complementary perspectives.
In addition, a key aspect of this work is the use of numerical methods to study the sensitivity of GW detectors to different polarizations while consistently incorporating realistic orbital effects.
We emphasize that no significant deviations from the standard assumptions, such as equal and constant arm lengths, are expected to affect our conclusions.

This paper is organized as follows.
In Sec.~\ref{sec:GW_signal}, we present the waveforms of binary black holes (BBHs) and SGWB with flat, power-law, and first-order PT spectra.
In Sec.~\ref{sec:Detector}, we review the relevant aspects of the detector, including the orbit configuration, response, TDI, noise, and sensitivity.
In Sec.~\ref{sec:Methodology}, we explain the approach for calculating the signal-to-noise ratio (SNR) and the procedural steps of the simulation.
In Sec.~\ref{sec:Results}, we present simulation results for different TDI combinations and analyze the best combinations under various scenarios.
Finally, we summarize the results of our research in Sec.~\ref{sec:Conclusion}.
Throughout this paper, we use units with
$G=c=1$, where $G$ is the gravitational constant and $c$ is the speed of light.

% =======================================
\section{GW signal}\label{sec:GW_signal}
In general, GWs have six polarizations: two tensor modes ($+$ and $\times$), two vector modes ($X$ and $Y$) and two scalar modes ($B$ and $L$)~\cite{ppE_Taiji}.
The GWs with six polarizations can be expressed in tensor form:
\begin{equation}
    h_{ij}=\sum_{A}h_A e^A_{ij},
\end{equation}
where $A=\{+,\times,X,Y,B,L\}$ represents the six polarizations, $h_A$ are the waveforms, and $e^A$ are the polarization tensors in the source frame.

\begin{figure}[ht]
    \begin{minipage}{\columnwidth}
        \centering
        \includegraphics[width=0.9\textwidth,
        trim=0 0 0 0,clip]{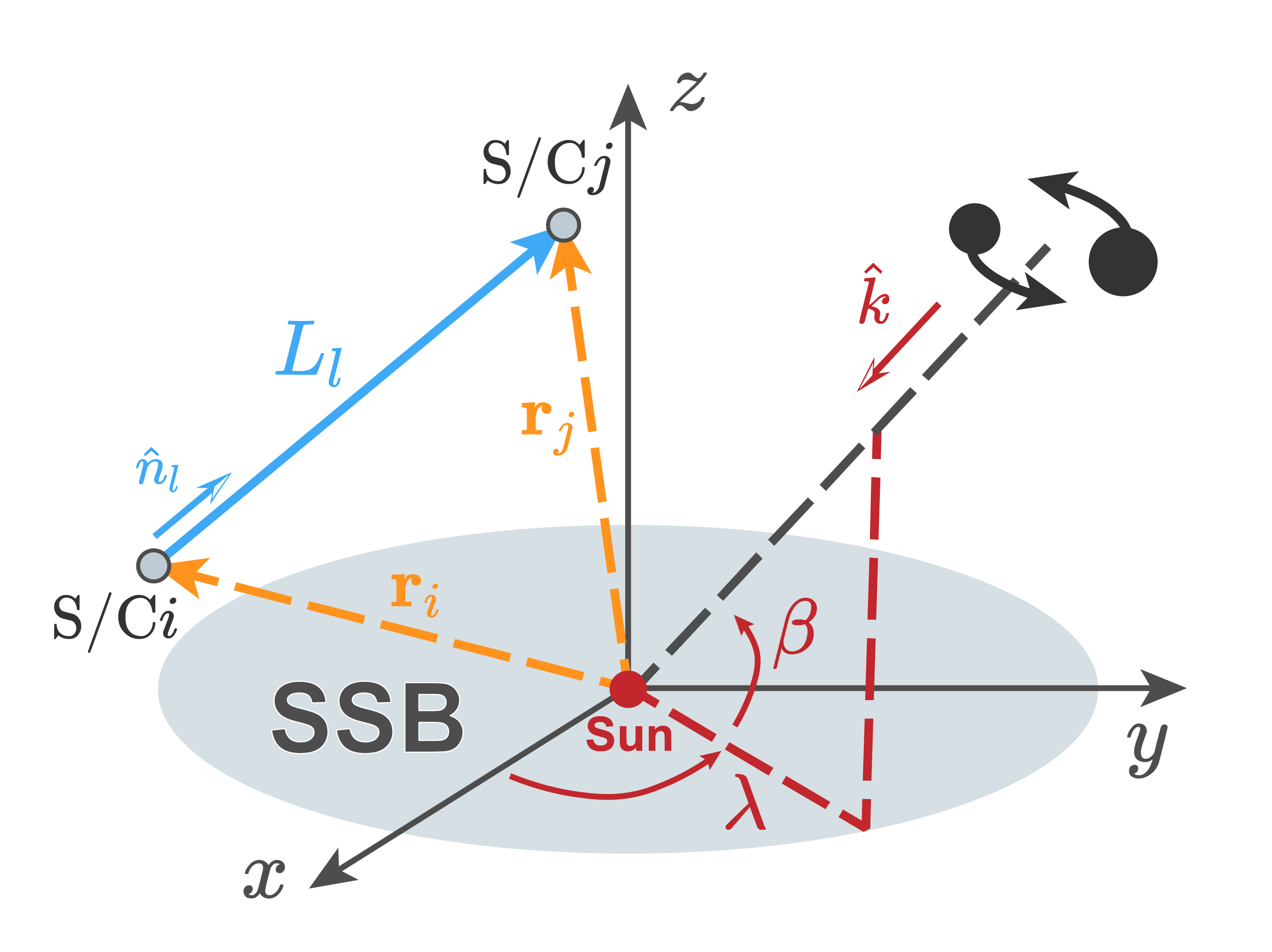}
        \caption{The diagram of the SSB frame. In the SSB frame, the $z$-axis is parallel to the orbital angular momentum of the Earth, while the $x$-axis points to the vernal equinox. }\label{fig:frame}
    \end{minipage}
\end{figure}

\begin{figure*}[ht] 
    \begin{minipage}{\textwidth}
        \centering
        \includegraphics[width=0.97\textwidth,
        trim=0 0 0 0,clip]{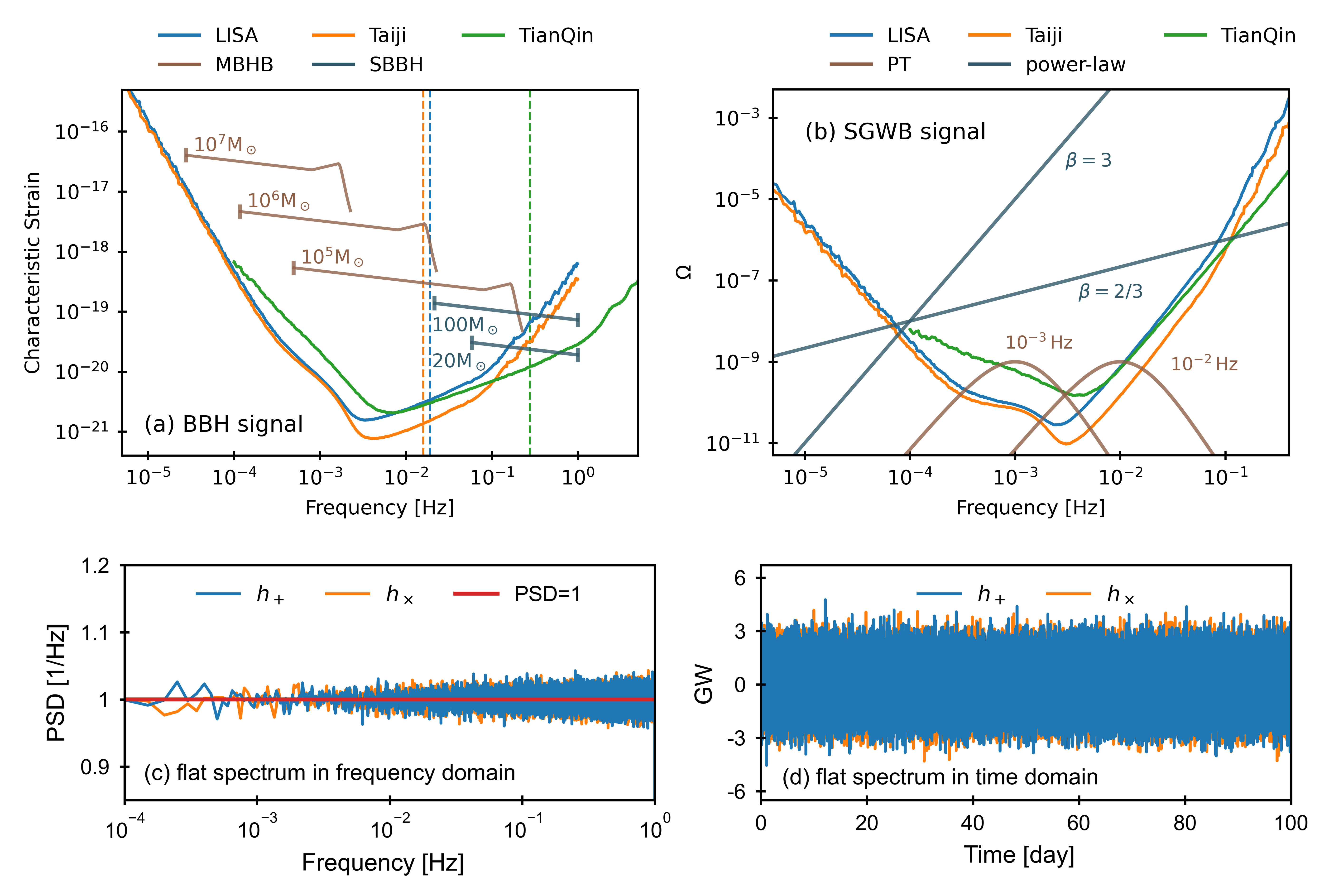}
        \caption{Comparison of different GW sources. (a) shows BBH signals and sensitivity curves, where the vertical axis represents the dimensionless characteristic strain $\sqrt{fS_n(f)}$. The BBH curve is derived from Ref.~\cite{LISA_noise} and is intended solely for illustrative purposes. For luminosity distance, we set SBBH at $D_L=44.6$ Mpc and MBHB at $D_L=6.79$ Gpc. The three vertical dashed lines represent the transfer frequencies $f_*$ of the three space-based detectors. (b) displays SGWB signals and the corresponding sensitivity curves, where the vertical axis denotes the dimensionless energy density spectrum $\Omega(f)$. For visualization purposes, the peak amplitude of the PT signal is fixed to $\Omega_{\rm p} = 10^{-9}$, and the amplitude and reference frequency of the power-law signal are set to $\Omega_{\beta} = 10^{-8}$ and $f_{\rm ref} = 0.1\,\mathrm{mHz}$. (c) and (d) illustrate a simulated SGWB source configured as white noise with $\mathrm{PSD}=1$, shown in the frequency domain and time domain, respectively. Only tensor modes are displayed, and the same setup applies to the other polarizations.}\label{fig:GW}
    \end{minipage}
\end{figure*}

According to Ref.~\cite{GWSpace}, we construct the solar system barycenter (SSB) frame, as shown in Fig.~\ref{fig:frame}. 
For GWs propagating along the $\hat{k}$, we can define the following right-handed orthonormal basis:
\begin{equation}
    \begin{aligned}
    &\hat{u}=[\sin\lambda, -\cos\lambda, 0] ,\\
    &\hat{v}=[-\sin\beta\cos\lambda, -\sin\beta\sin\lambda, \cos\beta] ,\\
    &\hat{k}=[-\cos\beta\cos\lambda, -\sin\lambda\cos\beta, -\sin\beta],
    \end{aligned}
\end{equation}
where $(\lambda, \beta)$ are ecliptic coordinates of the GW source.
The source frame can be transformed to the SSB frame via the polarization angle $\psi$. 
The polarization tensors $e^A$ in the source frame can be expressed using the polarization tensors $\epsilon ^A$ in the SSB frame~\cite{ppE_PTA,ppE_Taiji,GWSpace}:
\begin{equation}
    \begin{aligned}
    e^{+}&=\epsilon^{+}\cos2\psi+\epsilon^{\times}\sin2\psi,\\
    e^{\times}&=-\epsilon^{+}\sin2\psi+\epsilon^{\times}\cos2\psi,\\
    e^{X}&=\epsilon^{X}\cos\psi+\epsilon^{Y}\sin\psi,\\
    e^{Y}&=-\epsilon^{X}\sin\psi+\epsilon^{Y}\cos\psi,\\
    e^{B}&=\epsilon^{B},\\ e^{L}&=\epsilon^{L},
\end{aligned}
\end{equation}
with
\begin{equation}
    \begin{aligned}
    \epsilon^{+}&=\hat{u}\otimes\hat{u}-\hat{v}\otimes\hat{v} ,\\
    \epsilon^{\times}&=\hat{u}\otimes\hat{v}+\hat{v}\otimes\hat{u},\\
    \epsilon^{X}&=\hat{u}\otimes\hat{k}+\hat{k}\otimes\hat{u},\\
    \epsilon^{Y}&=\hat{v}\otimes\hat{k}+\hat{k}\otimes\hat{v},\\
    \epsilon ^{B}&=\hat{u}\otimes\hat{u}+\hat{v}\otimes\hat{v},\\
    \epsilon ^{L}&=\hat{k}\otimes\hat{k}.
    \end{aligned}
\end{equation}

For the BBH waveforms, the tensor modes $h_{+}$ and $h_{\times}$ are generated using the \texttt{SEOBNRv4} model~\cite{SEOBNR}. Since this model does not include waveforms with additional polarization modes, we adopt the parameterized post-Einsteinian (ppE) framework to describe and incorporate the non-tensorial polarizations.
Within this setup, we simulate typical-mass MBHBs and two SBBHs following our previous work~\cite{my_paper4}.
According to Refs.~\cite{ppE_PTA,ppE_Taiji,ppE2,ppE4}, the non-tensorial GW waveform under the ppE framework can be written as
\begin{equation}\label{eq:waveform}
\begin{aligned}
h_{X} & = \mathcal{A}_V \cos\iota\cos(\Phi+\Phi_0), 
\,
h_{Y} = \mathcal{A}_V \sin(\Phi+\Phi_0), \\
h_{B} & = \mathcal{A}_B \sin\iota\cos(\Phi+\Phi_0), 
\,
h_{L} = \mathcal{A}_L \sin\iota\cos(\Phi+\Phi_0),
\end{aligned}
\end{equation}
with
\begin{equation}\label{eq:amplitude}
    \mathcal{A}_{V,B,L} = \frac{\alpha_{V,B,L}}{D_L} \mathcal{M}^{4/3}\omega^{1/3},
\end{equation}
where $\alpha_{V,B,L}$ are the dimensionless ppE parameters, $\mathcal{M}=(m_1m_2)^{3/5}/(m_1+m_2)^{1/5}$ is the chirp mass, $m_1$ and $m_2$ are the masses of BBH, $D_L$ is the luminosity distance, $\iota$ is the inclination angle, $\Phi=\int{\omega}\mathrm{d} t$ is the orbital phase, $\omega$ is the orbital angular frequency, $\Phi_0$ is the initial orbital phase.
The overall evolution of orbital angular frequency can be described by considering the combined effects of dipole and quadrupole radiation~\cite{ppE_PTA,ppE_Taiji}
\begin{equation}\label{eq:angular_frequency}
    \frac{\mathrm{d}\omega}{\mathrm{d}t}=\alpha_D\eta^{2/5}\mathcal{M}\omega^3+\alpha_Q\mathcal{M}^{5/3}\omega^{11/3} ,
\end{equation}
where $\alpha_D$ and $\alpha_Q$ are the ppE parameters that describe the orbital angular frequency contributions of dipole and quadrupole radiation, $\eta=m_1m_2/M^2$ is the symmetric mass ratio, and $M=m_1+m_2$ is the total mass.
The analytical solution for the time evolution function $\omega(t)$ of the orbital angular frequency, as indicated by Eq.~(\ref{eq:angular_frequency}), poses a challenging task.
We determine $t(\omega)$ through integration,
\begin{equation}\label{eq:t(w)}
    t=t_0+\int_{\omega(t_0)}^{\omega(t)}{\left(\frac{\mathrm{d}\omega}{\mathrm{d}t}\right)^{-1}} \mathrm{d}\omega,
\end{equation}
which is shown in Refs.~\cite{ppE_PTA,ppE_Taiji}.
The bisection method and other computational techniques are employed to iteratively determine the orbital angular frequency corresponding to a given time point.
By solving point by point, we obtain the value of $\omega(t)$.
Using the above method, we input the calculated $\omega(t)$ into Eqs.~(\ref{eq:waveform})$\sim$(\ref{eq:amplitude}) to derive the final GW signal.
For MBHBs, the signal duration is set to 90 days. Specifically, the tensor modes include the complete waveform with merger and ringdown, while the additional polarization modes are terminated at the innermost stable circular orbit. 
For SBBHs, the duration of all polarization modes is set to one year prior to reaching 1 Hz.
Schematic illustrations of these waveforms in the frequency domain are shown in Fig.~\ref{fig:GW}(a).
Using the SEOBNRv4 model together with the ppE framework, we numerically construct the GW waveforms for different polarization modes.

For the SGWB, we consider two representative spectral shapes.
The first is the power-law spectrum, which is commonly written as~\cite{PL1,PL2}
\begin{equation}\label{eq:power-law}
\Omega_{\mathrm{GW}}(f) = \Omega_\beta \left( \frac{f}{f_{\mathrm{ref}}} \right)^{\beta},
\end{equation}
where $\beta$ denotes the spectral index and $f_{\mathrm{ref}}$ is a reference frequency.
Different values of $\beta$ correspond to distinct physical origins.
For instance, $\beta = 2/3$ is characteristic of astrophysical backgrounds from compact binary coalescences~\cite{PL_beta_23}, while $\beta = 0$ is often associated with slow roll inflation and
cosmic string models~\cite{PL_beta_0}.
The choice of $f_{\mathrm{ref}}$ is arbitrary and does not affect the detectability of the signal.
In this work, we set $f_{\mathrm{ref}} = 10^{-4}\,\mathrm{Hz}$.
Another important class of SGWB sources arises from first-order PTs in the early universe~\cite{PT1}.
GWs generated by phase boundary collisions in vacuum PTs exhibit a characteristic peaked spectrum~\cite{PT2,PT3}.
We adopt a phenomenological PT model obtained by fitting numerical simulations of PTs, given by~\cite{PT_fun1,PT_fun2}
\begin{equation}
\Omega_{\mathrm{PT}}(f) = \Omega_{\rm p}
\left( \frac{f}{f_{\rm pt}} \right)^3
\left( \frac{7}{4 + 3 (f / f_{\rm pt})^2} \right)^{7/2},
\end{equation}
where the peak frequency $f_{\rm pt}$ and the amplitude $\Omega_{\rm p}$ are determined by the underlying PT parameters, including the transition temperature, the Hubble rate, and the mean bubble separation.
Current theoretical models typically predict
$10^{-4}\,\mathrm{Hz} < f_{\rm pt} < 10^{-2}\,\mathrm{Hz}$ and
$10^{-14} < \Omega_{\rm p} < 10^{-9}$~\cite{PT_value,PT_value2}.
In our analysis, we fix $\Omega_{\rm p} = 10^{-9}$ for illustration.

Since our primary goal is to compare the relative performance of different TDI combinations,
the specific normalization constants adopted in these GW models do not affect the final conclusions.
The relation between the dimensionless energy density spectrum $\Omega(f)$
and the one-sided power spectral density (PSD) $S(f)$ is given by~\cite{PT_value}
\begin{equation}
\Omega(f) = \frac{4\pi^2}{3 H_0^2} f^3 S(f),
\end{equation}
where the Hubble constant is taken as
$H_0 = 67.4\,\mathrm{km\,s^{-1}\,Mpc^{-1}} \simeq 2.2 \times 10^{-18}\,\mathrm{Hz}$.
Figure~\ref{fig:GW}(b) illustrates the characteristic features of the power-law and PT signals in terms of the energy density spectrum.
The sensitivity curves shown in this panel are obtained by converting those presented in Fig.~\ref{fig:GW}(a) into the corresponding energy density representation.

For the purpose of calculating the GW average response and sensitivity curves, a flat PSD of the SGWB is introduced here. 
Such an isotropic flat SGWB is equivalent to white noise covering all frequencies at once~\cite{LISA_Sensitivity}. 
As depicted in Figs.~\ref{fig:GW}(c) and (d), the time-domain signal is generated by employing the PSD in the frequency domain and adding random phase information. 
In the subsequent computations, the SGWB corresponding to the six polarizations are all generated.
We will elaborate on this in detail in Sec.~\ref{subsec:Sensitivity}.
For a more concise presentation of the results, we primarily consider the tensor and vector modes. 
Specifically, the tensor mode consists of both $h_+$ and $h_\times$, while the vector mode includes both $h_X$ and $h_Y$. 
This selection allows for a clearer comparison of different polarization sensitivities in our analysis.

% =======================================
\section{Detector}\label{sec:Detector}
\subsection{Detector orbit and response}\label{subsec:Detector_orbit_and_response}
The millihertz space-based GW detectors, LISA, Taiji, and TianQin, are all composed of three S/C, forming a nearly equilateral triangle. 
LISA and Taiji adopt heliocentric orbits, trailing/leading the Earth by approximately 20 degrees with a period of one year~\cite{LISA,Taiji}.
TianQin employs a geocentric orbit, and the normal direction of the detector plane always points towards the reference source RX J0806.3+1527~\cite{TianQin}. 
The arm lengths $L$ of the detectors are $2.5\times10^6$ km (LISA), $3\times10^6$ km (Taiji), and $\sqrt{3}\times10^5$ km (TianQin), respectively.

The motion of the S/C can be described by the Keplerian orbit. 
In the SSB frame, the position vector of the $i$-th S/C is $\mathbf{r}_i =(x_i,y_i,z_i)$.
For the heliocentric and geocentric orbits, the analytical components of $\mathbf{r}_i$ as a function of time are provided in Refs.~\cite{Taiji_orbit,TianQin_orbit}. 
We employ the Keplerian orbit expanded to the order of $e^2$, where $e$ is the orbital eccentricity. 

\begin{figure}[ht]
    \begin{minipage}{\columnwidth}
        \centering
        \includegraphics[width=0.8\textwidth,
        trim=0 0 0 0,clip]{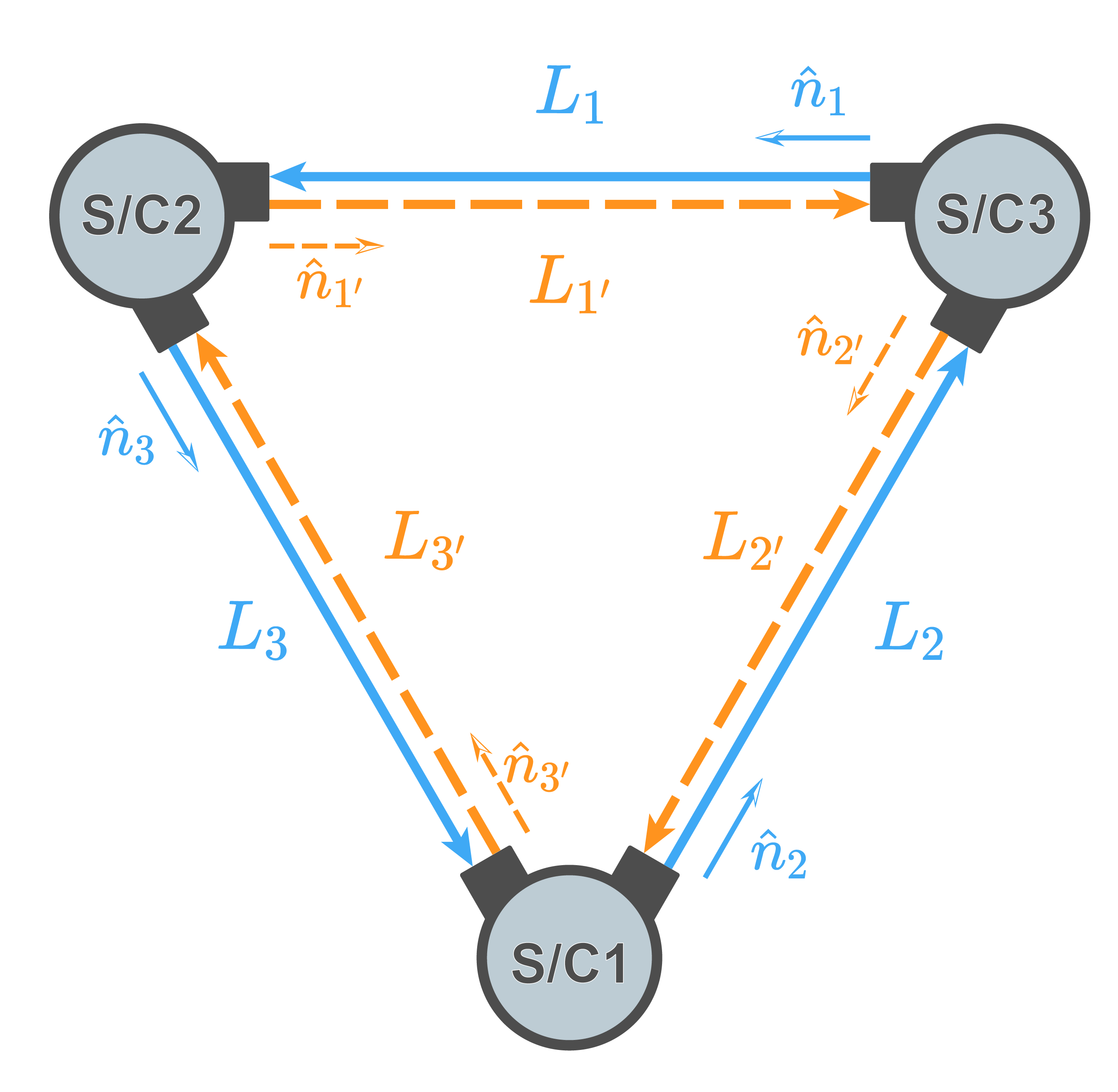}
        \caption{The constellation of space-based detector. }\label{fig:three_SC}
    \end{minipage}
\end{figure}

The Michelson interferometer detects GWs by measuring the relative changes in the lengths of its two arms. 
The change in length of one arm induced by a GW signal is given by~\cite{GWSpace}
\begin{equation}
    H(t)=n_l^i h_{ij}(t)n_l^j,
\end{equation}
where $\hat{n}_l$ is the unit vector of the photon propagation.
Following Refs.~\cite{GWSpace,Taiji_data_challenge}, the single arm response to GW between $\mathrm{S/C}i$ to $\mathrm{S/C}j$ in time domain can be defined as
\begin{equation}
    \eta^{\mathrm{GW}}=\frac{1}{2(1-\hat{k}\cdot\hat{n}_{l})}\left[H(t-\hat{k}\cdot\mathbf{r}_{i}-L_l)-H(t-\hat{k}\cdot\mathbf{r}_{j})\right],
\end{equation}
where $L_l$ is the distance of the photon propagation between $\mathrm{S/C}i$ to $\mathrm{S/C}j$.
Some markers and definitions are shown in Figs.~\ref{fig:frame} and \ref{fig:three_SC}.
Considering the propagation time of light, the round-trip distances from $\mathrm{S/C}i$ to $\mathrm{S/C}j$ are not equal, $L_l\ne L_{l'}\ne L$, which are defined as~\cite{PhD}
\begin{equation}\label{eq:L}
    \begin{aligned}
    L_1(t)&=\left|\mathbf{r}_2(t-L)- \mathbf{r}_3(t)\right|, \\
    L_{1'}(t)&=\left|\mathbf{r}_3(t-L)- \mathbf{r}_2(t)\right|.\\
    \end{aligned}
\end{equation}
All other $L_l$ can be derived from the cyclical indicators.
The unit vector $\hat{n}_l$ is also the same, with $\hat{n}_l\ne-\hat{n}_{l'}$, calculated using the method in Eq.~(\ref{eq:L}).

% ---------------------------------------
\subsection{Time-delay interferometry}\label{subsec:Time-delay_interferometry}
TDI is a kind of data processing technique, which combines the measurements conducted at different times to realize noise suppression~\cite{TDI_LRR}.
The TDI technique constructs equivalent equal-arm interferometry by combining multiple-link measurements, which can be divided into several generations~\cite{Taiji_data_challenge}.
For the first-generation TDI, in the stationary situation, the arm length is constant in time, with $L_i=L_{i'}=\mathrm{const}$. 
The modified first-generation TDI, also referred to as TDI-1.5, employ the rigid rotating case, with $L_i=\mathrm{const}$, $L_{i'}=\mathrm{const}$, $L_i\ne L_{i'}$. 
In the second-generation TDI, a flexing case is adopted, and the arm length is a function of time, with $L_i=L_i(t)$, $L_{i'}=L_{i'}(t)$, $L_i\ne L_{i'}$. 
In this paper, we adopt the second-generation TDI and conform to the actual situation (see Eq.~(\ref{eq:L})). 

For an arbitrary time-dependent variable $\eta(t)$, a time-delay operator $D_i$ can be defined as
\begin{equation}\label{eq:D}
    \begin{aligned}
    D_i\eta(t)&=\eta(t-L_i(t)), \\
    D_{ji}\eta(t)&=\eta(t-L_j(t)-L_i(t-L_j(t))).\\
    \end{aligned}
\end{equation}
Then, we can construct various TDI combinations~\cite{TDI_combination1}. 
Here are three typical TDI combinations~\cite{TDI_combination2}: 
the Michelson combination ($X,Y,Z$),
the Sagnac combination ($\alpha, \beta, \gamma $),
and the fully symmetric combination ($\zeta $)
An important issue is the failure of one laser link, such as a laser source failure. 
In such case, several TDI combinations that rely on single-link measurements can be used to reconstruct the GW signal~\cite{PhD}.
These are called link failure surviving combinations~\cite{TDI_different}:
the Beacon combination ($P, Q, R$), 
the Monitor combination ($E, F, G$),
and the Relay combination ($U, V, W$).
The specific forms of these different TDI are described in detail in Appendix~\ref{sec:TDI_conbinations}.

By linearly combining the $X,Y,Z$ channels of the Michelson combination, the so-called optimal combination ($\mathcal{A},\mathcal{E},\mathcal{T}$) can be obtained, defined as~\cite{TDI_AET}
\begin{equation}\label{eq:AET}
    \mathcal{A}=\frac{Z-X}{\sqrt2} ,\ \mathcal{E}=\frac{X-2Y+Z}{\sqrt6} ,\ \mathcal{T}=\frac{X+Y+Z}{\sqrt3} .
\end{equation}
In an ideal situation, with equal arms and equal noise levels, this combination can eliminate the mutual correlation between signals and noise in different channels~\cite{TDI_AET}. 
The $\mathcal{T}$ channel is also known as the null channel, which has poor sensitivity in the low-frequency range.
Unlike other TDI combinations, this particular combination renders the three channels asymmetric, thereby distinguishing the multi-channel scenario. 
This distinction is elaborated in Sec~\ref{subsec:AET_channels}.

% ---------------------------------------
\subsection{Noise and sensitivity}\label{subsec:Sensitivity}
In the payload of a space-based detector, there are two optical platforms on each S/C, and each platform outputs three data streams, which contain both GW signals and various noise.
By combining these data streams, the application of TDI yields the final datasets corresponding to different channels.
With the implementation of TDI techniques together with sideband modulation generated by electro-optical modulators, laser frequency noise and clock noise can be suppressed below the instrumental noise level~\cite{clock_noise1,clock_noise2}.  
As a result, we focus primarily on the instrumental noise contributions of the detector. 

\begin{figure*}[ht] 
    \begin{minipage}{\textwidth}
        \centering
        \includegraphics[width=0.97\textwidth,
        trim=0 0 0 0,clip]{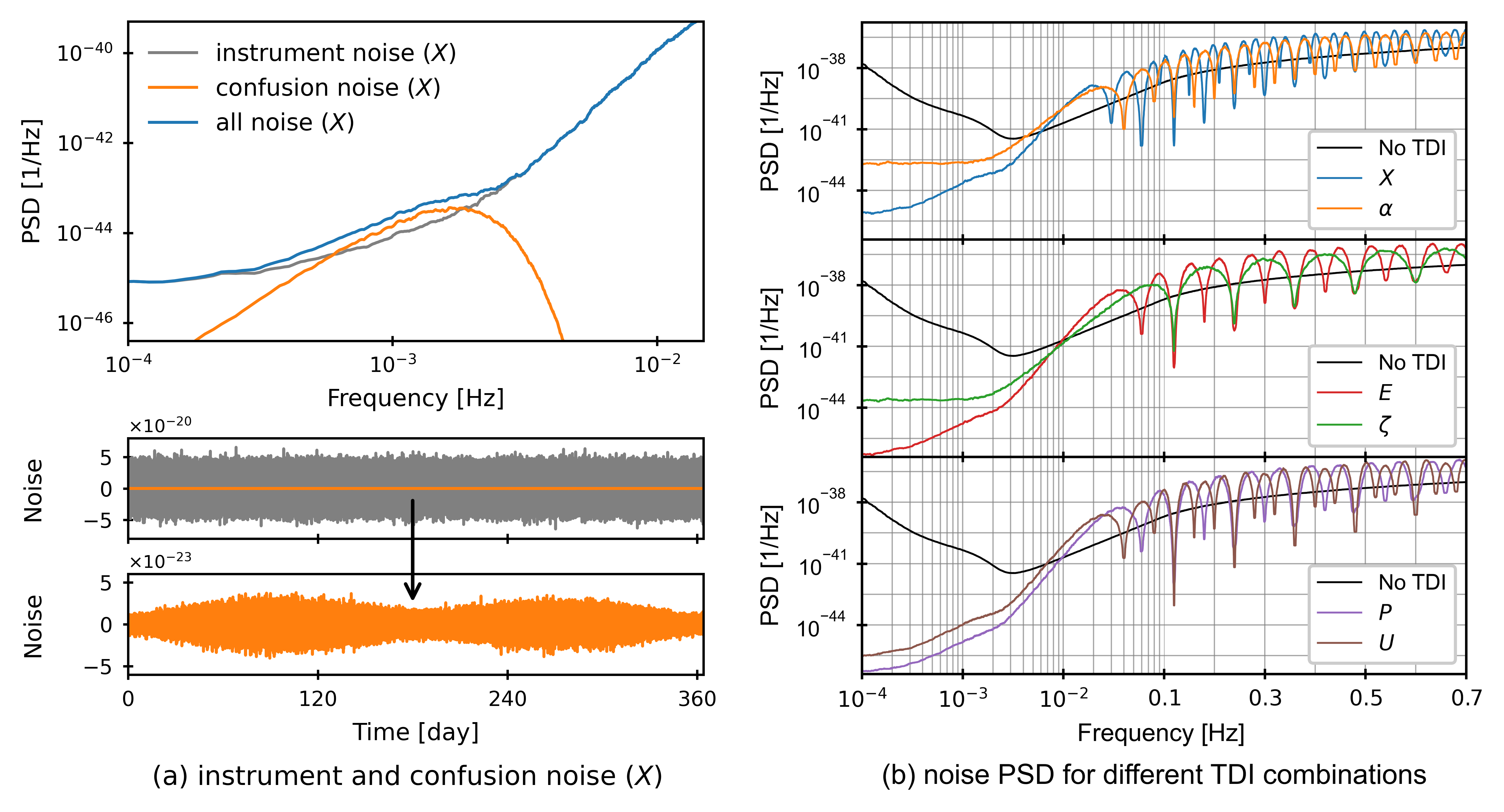}
        \caption{Comparison of the noise PSD for LISA. (a) shows the instrumental noise and the confusion noise in the $X$ channel. The upper subpanel presents the noise PSD in the frequency domain, while the two lower subpanels illustrate the corresponding signals transformed into the time domain. (b) shows the noise PSD for different second-generation TDI combinations. Frequencies below 0.1 Hz are shown on a logarithmic scale. \textit{No TDI} refers to the interference only between adjacent links, such as $\eta_{31}-\eta_{21}$. Both instrumental noise and confusion noise are included. All sensitivity curves and PSD in this paper are numerically simulated. For clarity and conciseness, median smoothing is used.}\label{fig:noise}
    \end{minipage}
\end{figure*}

The instrumental noise of the detector consists of acceleration noise $n^{\mathrm{TM} }$ and optical metrology noise $n^{\mathrm{OMS} }$~\cite{TDI_noise1,TDI_noise2}.
Neglecting laser frequency noise and clock noise, the noise terms composed of different data streams can be written as~\cite{TDI_eta1,TDI_eta2}
\begin{equation}
    \eta^{\mathrm{Noise} }(t)=n^{\mathrm{OMS} }_l(t)+D_l n^{\mathrm{TM} }_{l'}(t)+n^{\mathrm{TM} }_{l}(t).
\end{equation}
Assuming that the instrumental noise is Gaussian and stationary, the time-domain noise data are generated by transforming the frequency-domain noise PSD and adding random phases.
For LISA and Taiji, the instrumental noise is modeled as the combination of optical metrology noise $S^{\mathrm{OMS}}(f)$ and test-mass acceleration noise $S^{\mathrm{TM}}(f)$, given by~\cite{LISA_noise,Taiji_data_challenge} 
\begin{equation}
S^{\mathrm{OMS}} = S_x \left( \frac{2\pi f}{c} \right)^2 
\left[ 1 + \left( \frac{2\,\mathrm{mHz}}{f} \right)^4 \right],
\end{equation}
\begin{equation}
S^{\mathrm{TM}} = \frac{S_a}{(2\pi f c)^2}
\left[ 1 + \left( \frac{0.4\,\mathrm{mHz}}{f} \right)^2 \right]
\left[ 1 + \left( \frac{f}{8\,\mathrm{mHz}} \right) \right].
\end{equation}
We adopt $\sqrt{S_x} = 15\,\mathrm{pm\,Hz^{-1/2}}$ for LISA and
$\sqrt{S_x} = 8\,\mathrm{pm\,Hz^{-1/2}}$ for Taiji, while both detectors share
$\sqrt{S_a} = 3 \times 10^{-15}\,\mathrm{m\,s^{-2}\,Hz^{-1/2}}$.
For TianQin, the instrumental noise is described by~\cite{GWSpace}
\begin{equation}
S^{\mathrm{OMS}} = S_x \left( \frac{2\pi f}{c} \right)^2,
\end{equation}
\begin{equation}
S^{\mathrm{TM}} = \frac{S_a}{(2\pi f c)^2}
\left[ 1 + \left( \frac{0.1\,\mathrm{mHz}}{f} \right) \right],
\end{equation}
with $\sqrt{S_x} = 1\,\mathrm{pm\,Hz^{-1/2}}$ and
$\sqrt{S_a} =1\times 10^{-15}\,\mathrm{m\,s^{-2}\,Hz^{-1/2}}$.
The data in a single arm are then constructed by independently generating the noise contribution
$\eta^{\mathrm{Noise}}(t)$ and the GW signal $\eta^{\mathrm{GW}}(t)$ for each link.
By employing different TDI combinations, the data $X^{\mathrm{Noise} }(t)$ and $X^{\mathrm{GW} }(t)$ in different channels can be acquired. 
Such a method can be used to simulate the scenarios of different TDI combinations.

In addition to the instrumental noise, an important noise source for space-based detectors arises from the superposition of GWs emitted by millions of galactic binaries, forming the so-called galactic foreground or confusion noise~\cite{GB1,GB2,GB3}.
In the frequency range $0.5-3\,\mathrm{mHz}$, this confusion noise can exceed the instrumental noise and thus affect the detector sensitivity.
By simulating galactic binaries in the Milky Way, we iteratively identify and subtract high-SNR resolvable sources. % 改
The remaining unresolved population forms the galactic confusion noise, which is then projected into different TDI combinations through the sky-averaged detector response. % 改
Specifically, we model the confusion noise $S^{\mathrm{GB}}(f)$ for different detectors using a polynomial fitting function in logarithmic scale, expressed as
\begin{equation}
S^{\mathrm{GB}} = 10^{x},
\end{equation}
with
\begin{equation}
x = \sum_{n=0}^{5} a_n
\left[
\log_{10}\left( \frac{f}{1\,\mathrm{mHz}} \right)
\right]^n ,
\end{equation}
where the coefficients $a_n$ for the three detectors can be found in Ref.~\cite{my_paper2}.
Since galactic binaries are mainly concentrated in the bulge region of the Galaxy, the confusion noise is modulated by the detector motion.
Therefore, the time-domain confusion noise is constructed by transforming the frequency-domain realization into the time domain and multiplying it by a time-varying multiplier, which can be written as~\cite{PT_fun2}
\begin{equation}
S^{\mathrm{GB}}(f,t)=S^{\mathrm{GB}}(f)\left[
1+\sum_{n}A_n \cos(\omega_n t + \phi_n) \right],
\end{equation}
where $\omega_n = 2\pi n / (1\,\mathrm{yr})$ encodes the annual modulation effect. 
The corresponding Fourier coefficients $A_n$ and phases $\phi_n$ are adopted from Ref.~\cite{GB_time}. % 改
Given that current GW observations show no evidence for deviations from GR, any additional polarization components emitted by galactic binaries, if present, are expected to be much weaker than the tensor modes~\cite{test_GR1,test_GR2,test_GR3}.
As a result, the confusion noise has a negligible impact on the sensitivity curves for additional polarizations.
We therefore restrict the effect of confusion noise to the sensitivity curves of tensor modes only.

Figure~\ref{fig:noise}(a) illustrates the confusion noise after the $X$-channel response in both the frequency and time domains.
A pronounced bump appears in the frequency domain around $1\,\mathrm{mHz}$, where the confusion noise exceeds the instrumental noise and can significantly affect GW signals in this frequency band. % 改
When examined in the time domain, however, the total noise budget receives contributions from the entire frequency range. % 改
As a result, the overall level of the confusion noise in the time domain remains several orders of magnitude lower than that of the instrumental noise. % 改

Figure~\ref{fig:noise}(b) further presents the noises processed by different second-generation TDI combinations, where distinct TDI choices lead to different final noise performances.
Simply focusing on the PSD of the noise is not sufficient for evaluation, as the GW response is also different for different TDI combinations.
In particular, the TDI transfer functions reshape the noise spectra and can suppress the effective noise level at low frequencies, while the same transfer functions also modify the GW signal response. % 改
One form of evaluating the detector performance is the sensitivity curve. 
The computation of TDI relies on the polarization angle and sky position of the source, and the sensitivity is calculated using the average of the responses to these parameters. 
It is typically accomplished through numerical or semi-analytical methods.
In this paper, a numerical method is employed to compute the sensitivity curves~\cite{LISA_Sensitivity}.

The response criterion is based on white noise covering the entire frequency band, equivalent to the flat SGWB described in Sec.~\ref{sec:GW_signal}. 
To compute the average response, the polarization angles and sky positions of one hundred sources are uniformly sampled using random sampling. 
The average response is written as:
\begin{equation}
    R_A=\left \langle \tilde{X}^\mathrm{GW} (f)\right \rangle ,
\end{equation}
where $<\cdot >$ represents the average value.
This ensures that the sky and the polarization angle spaces resolution is $(4\pi^2\times 2\pi)/100$. 
Our calculations indicate that when more than 20 random sources are used, the averaged results become stable, ensuring the reliability of our choice of 100 sources.
Therefore, the sensitivity can be defined as
\begin{equation}
    \sqrt{S_n}=\sqrt{P_n/R_A} ,
\end{equation}
where $P_n$ is the PSD of noise.
By this approach, the sensitivity curves of different TDI combinations are computed, and they are utilized subsequently for evaluating the performance of the detector and calculating the SNR of different polarizations.

\begin{figure}[ht]
    \begin{minipage}{\columnwidth}
        \centering
        \includegraphics[width=0.98\textwidth,
        trim=0 0 0 0,clip]{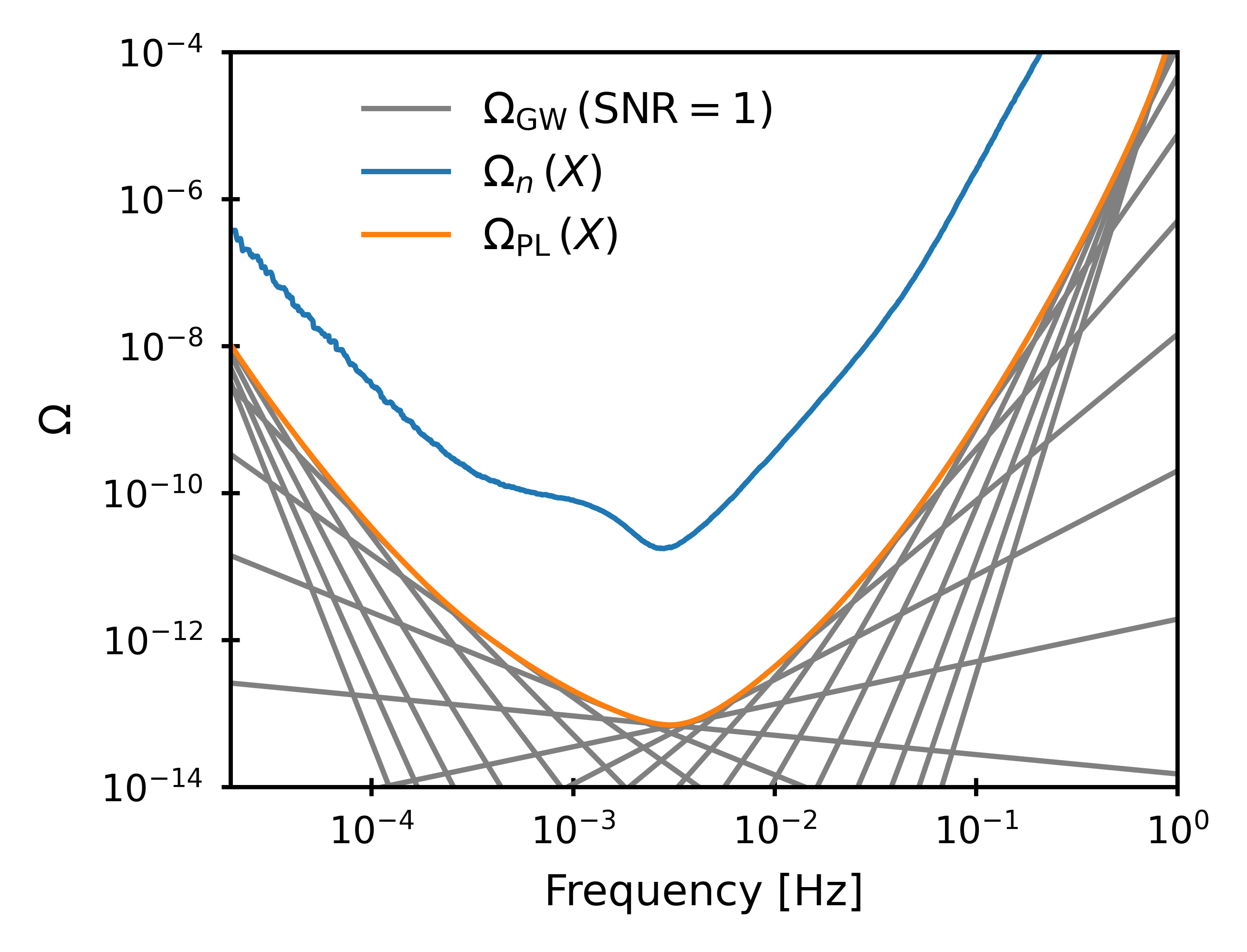}
        \caption{Computation of the PLIS curve for an observation time of one year and $\mathrm{SNR}=1$ for LISA. The gray curves correspond to power-law SGWB spectra with different spectral indices with $\mathrm{SNR}=1$. The LISA's sensitivity curve is also shown for reference.
}\label{fig:SGWB}
    \end{minipage}
\end{figure}

Based on the sensitivity curves, the SNR of a given GW signal can be efficiently evaluated.
To further enable a rapid assessment of power-law SGWB detectability, we introduce the power-law integrated sensitivity (PLIS).
The PLIS characterizes the minimum detectable amplitude of a family of power-law SGWB spectra for a given observation time and a threshold SNR.
It therefore provides a convenient tool to assess the detectability of SGWBs with different spectral indices and to directly compare the relative sensitivities of different detectors or TDI combinations.
For a set of spectral indices, the amplitude $\Omega_\beta$ corresponding to a given SNR can be computed as~\cite{PLIS}
\begin{equation}
\Omega_\beta =
\frac{\rho}{\sqrt{T}}
\left[
\int_{0}^{\infty}
\frac{(f/f_{\mathrm{ref}})^{2\beta}}{\Omega_n^2(f)} \, \mathrm{d}f
\right]^{-1/2}.
\end{equation}
In this work, we adopt SNR $\rho = 1$ and an observation time of $T = 1\,\mathrm{year}$ for PLIS calculation.
This choice allows the PLIS to be easily rescaled to other SNR thresholds or observation durations.

The envelope formed by the set of power-law spectra with different $\beta$ and $\Omega_\beta$ defines the PLIS curve, which can be written as
\begin{equation}
\Omega_{\mathrm{PL}} =
\underset{\beta}{\mathrm{max}}
\left[
\Omega_\beta \left( \frac{f}{f_{\mathrm{ref}}} \right)^{\beta}
\right].
\end{equation}
In Fig.~\ref{fig:SGWB}, the PLIS curve is constructed from a set of power-law spectra with spectral indices in the range $\beta = -8$ to $8$.
In logarithmic space, any power-law spectrum tangent to the PLIS curve corresponds to an integrated SNR equal to one.
Consequently, if a predicted SGWB spectrum lies entirely below the PLIS curve, its SNR is smaller than one.
Further details can be found in Ref.~\cite{PLIS}.

In summary, this section presents a comprehensive description of the noise modeling and sensitivity construction for space-based GW detectors.
We introduce the main instrumental noise sources, including the optical metrology noise and the test-mass acceleration noise, as well as the galactic confusion noise arising from unresolved galactic binaries.
By constructing the data streams, we compute the noise PSDs for different TDI combinations.
Using a flat SGWB, we further evaluate the averaged detector response and derive the corresponding sensitivity curves.
In particular, we introduce the PLIS as an efficient tool to assess the detectability of power-law SGWB spectra.
Based on these sensitivity curves, the SNRs of BBH signals and PT-induced SGWBs for different detectors and TDI combinations will be quantitatively evaluated in the following sections.

% =======================================
\section{Methodology}\label{sec:Methodology}
In general, the BBH SNR $\rho_A$ of GW polarization $h_A$ can be defined as
\begin{equation}
    \rho_A^2=(h_A|h_A),
\end{equation}
where the inner product $(\cdot|\cdot)$ generalizes the time-domain correlation product and is conventionally defined as
\begin{equation}
    (a|b)=4\text{Re}\left[\int_0^{\infty}\frac{\tilde{a}^*(f)\tilde{b}(f)}{S_n(f)}\mathrm{d}f\right],
\end{equation}
where $\tilde{a}(f)$ and $\tilde{b}(f)$ are the Fourier transforms of $a(t)$ and $b(t)$, respectively.
In this paper, different polarizations are considered to be independent~\cite{my_paper4}.
The overall SNR can be calculated by taking the inner product sum of the SNRs for different polarizations, and it can be expressed as
\begin{equation}\label{eq:SNR_all}
    \rho^2=\sum_{A}{\rho_A^2} =\sum_{A}{(h_A|h_A)}.
\end{equation}
For the SGWB, we consider an idealized autocorrelation measurement.
The corresponding SNR can be expressed as~\cite{LISA_Sensitivity,PT_value}
\begin{equation}
\rho^2 = T \int_{0}^{\infty}
\frac{\Omega^2_{\mathrm{GW}}(f)}{\Omega^2_n(f)} \, \mathrm{d}f ,
\end{equation}
where $\Omega_n(f)$ is the sensitivity $S_n(f)$ in energy density spectrum.

\begin{figure}[ht]
    \begin{minipage}{\columnwidth}
        \centering
        \includegraphics[width=0.98\textwidth,
        trim=0 0 0 0,clip]{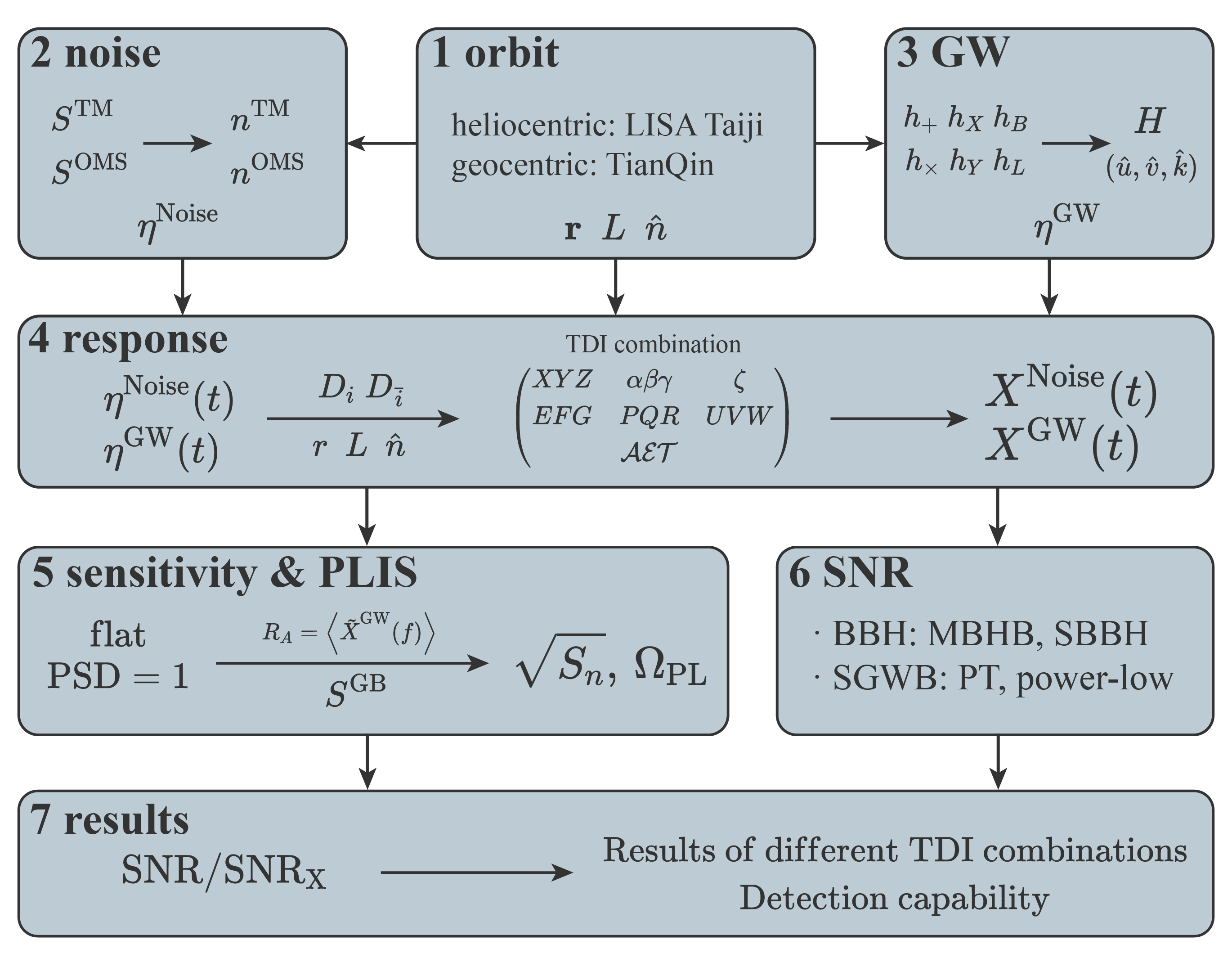}
        \caption{The flowchart of the simulation and calculation process in this paper. }\label{fig:process}
    \end{minipage}
\end{figure}

Based on such a computational method, we conduct calculations on the simulated data.
The simulation and calculation process of this paper is shown in Fig.~\ref{fig:process}, which can be summarized as follows:
\begin{enumerate}
    \item \textit{orbit}: By considering the orbital motions of LISA, Taiji, and TianQin separately, we calculate the light propagation path and direction between the S/C (Sec.~\ref{subsec:Detector_orbit_and_response}).

    \item \textit{noise}: We compute the time-domain instrumental noise from the frequency-domain noise, include the confusion noise, and construct the single-arm noise data stream $\eta^{\mathrm{Noise}}$ (Sec.~\ref{subsec:Sensitivity}). 

    \item \textit{GW}: We calculate GW signals from different polarizations, including those from BBH and SGWB, and construct the single-arm GW data stream $\eta^{\mathrm{GW}}$ (Secs.~\ref{sec:GW_signal} and \ref{subsec:Detector_orbit_and_response}).

    \item \textit{response}: Using various TDI combinations, we simulate the data for both noise and GW signals after TDI processing (Sec.~\ref{subsec:Time-delay_interferometry}).

    \item \textit{sensitivity \& PLIS}: Through numerical computation, we calculate the average response and derive both the sensitivity curves and the PLIS curves for different TDI combinations (Sec.~\ref{subsec:Sensitivity}). 

    \item \textit{SNR}: We compute the SNRs of BBH signals in the ppE framework and of SGWB signals for different polarizations (Sec.~\ref{sec:Methodology}). 

    \item \textit{results}: We compare the performance of different TDI combinations, evaluating the detection capabilities of different detectors (Sec.~\ref{sec:Results}).
\end{enumerate}

In summary, we outline the step-by-step methodology used to simulate and analyze the performance of space-based GW detectors.
Next, we present the results of these simulations, comparing the performance of different TDI combinations and analyzing their implications for GW detection.

% =======================================
\section{Results}\label{sec:Results}
\subsection{Constraints on polarization}\label{subsec:Constraints_on_polarization}

\begin{figure*}[ht]
    \begin{minipage}{\textwidth}
        \centering
        \includegraphics[width=0.97\textwidth,
        trim=0 0 0 0,clip]{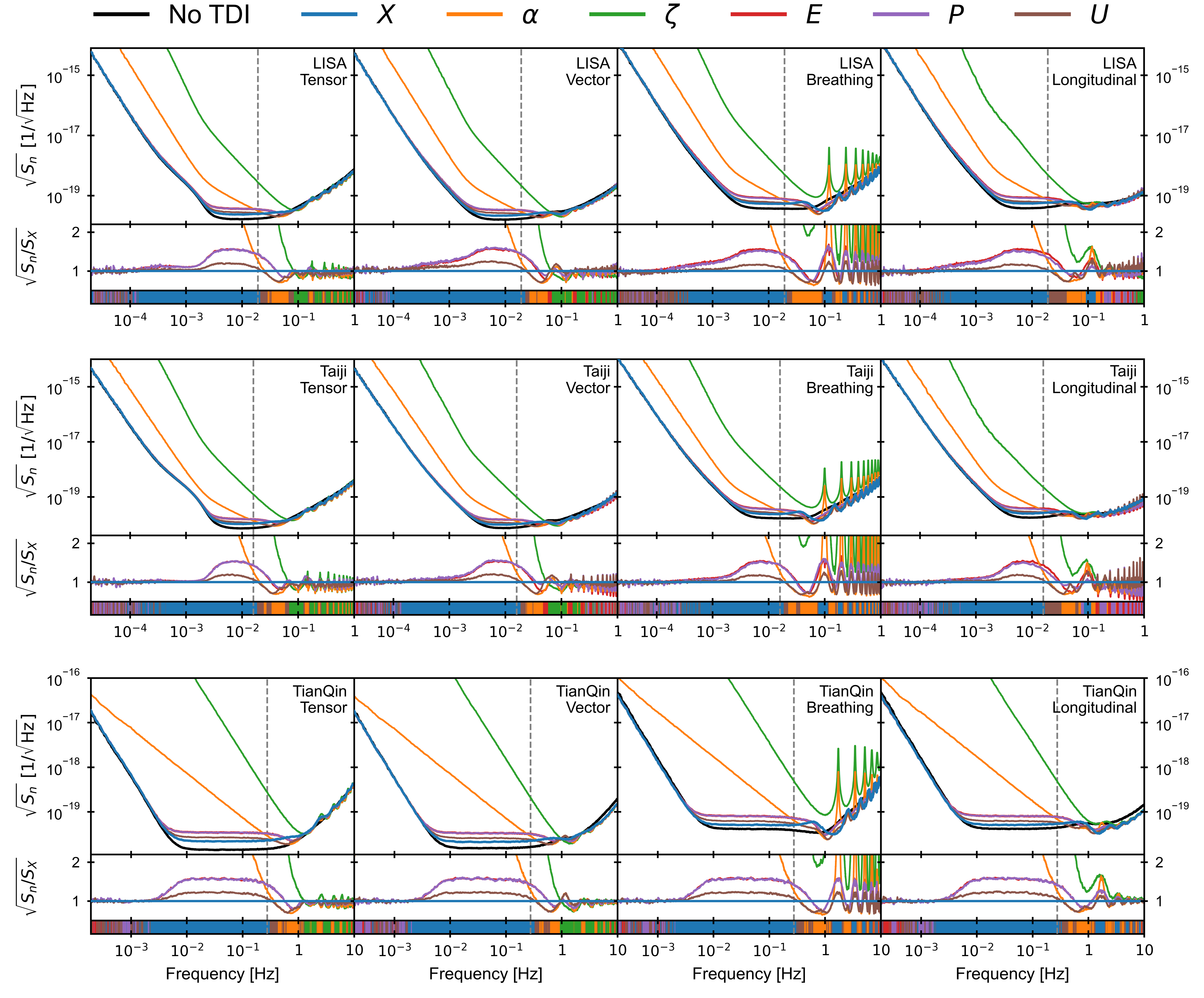}
        \caption{Comparison of sensitivity curves for various polarizations under different TDI combinations. The sensitivity curves of LISA, Taiji, and TianQin are calculated separately. Only the tensor modes include the effect of confusion noise, while the additional polarization modes are shown without confusion noise. The second row of each small graph represents the ratio relative to the $X$ channel as the benchmark. The color bar at the bottom represents the TDI combination with the best sensitivity at that frequency. The vertical-dashed line represents transfer frequency $f_* = 1/(2\pi L)$~\cite{transfer_function}. }\label{fig:sensitivity}
    \end{minipage}
\end{figure*}

We simulate and calculate the sensitivity curves in different TDI combinations for different polarizations. 
The results of LISA, Taiji, and TianQin are shown in Fig.~\ref{fig:sensitivity}.

In general, the lower the sensitivity curve, the better the sensitivity and the stronger the performance of the detector.
In the lower frequency range, the sensitivity of the $\alpha$ channel is better than that of the $\zeta$ channel. 
The results of other TDI combinations are regionally consistent and all lower than the former two. 
In the higher frequency range, the sensitivity curves of the $\alpha$ and $\zeta$ channels in the breathing mode present different characteristics. 
The differences in other TDI combinations or other polarizations are not significant.

In the most sensitive frequency band, the sensitivity curves of different TDI combinations exhibit distinct behaviors. 
The $X$ channel is the best, followed by the $U$ channel. 
The $E$ and $P$ channels are nearly the same, both inferior to the $X$ and $U$ channels, and the results of the $\alpha$ and $\zeta$ channels are the poorest. 
Compared with the sensitivity curve without TDI processing, the sensitivity curve after different TDI combinations has a loss in the most sensitive band, although the difference is not significant in the lower and higher frequency bands.

For space-based detectors, the transfer frequency $f_*$ serves as an important dividing line. 
When the frequency is lower than $f_*$, the $X$ channel possesses nearly the best sensitivity. 
When the frequency is higher than $f_*$, the sensitivity curves of different TDI combinations interlace with each other, and the best TDI combination varies in different frequency bands. 
Thus, it can be roughly stated that when the GW signal is in the frequency band lower than $f_*$, selecting the $X$ channel is the best choice. 
When it is in the frequency band higher than $f_*$, the TDI combination should be flexibly chosen in accordance with the characteristics of the signal and the requirements of the analysis.
For single-arm damage cases, the $U$ channel has better sensitivity compared with other channels.

\begin{figure*}[ht]
    \begin{minipage}{\textwidth}
        \centering
        \includegraphics[width=0.97\textwidth,
        trim=0 0 0 0,clip]{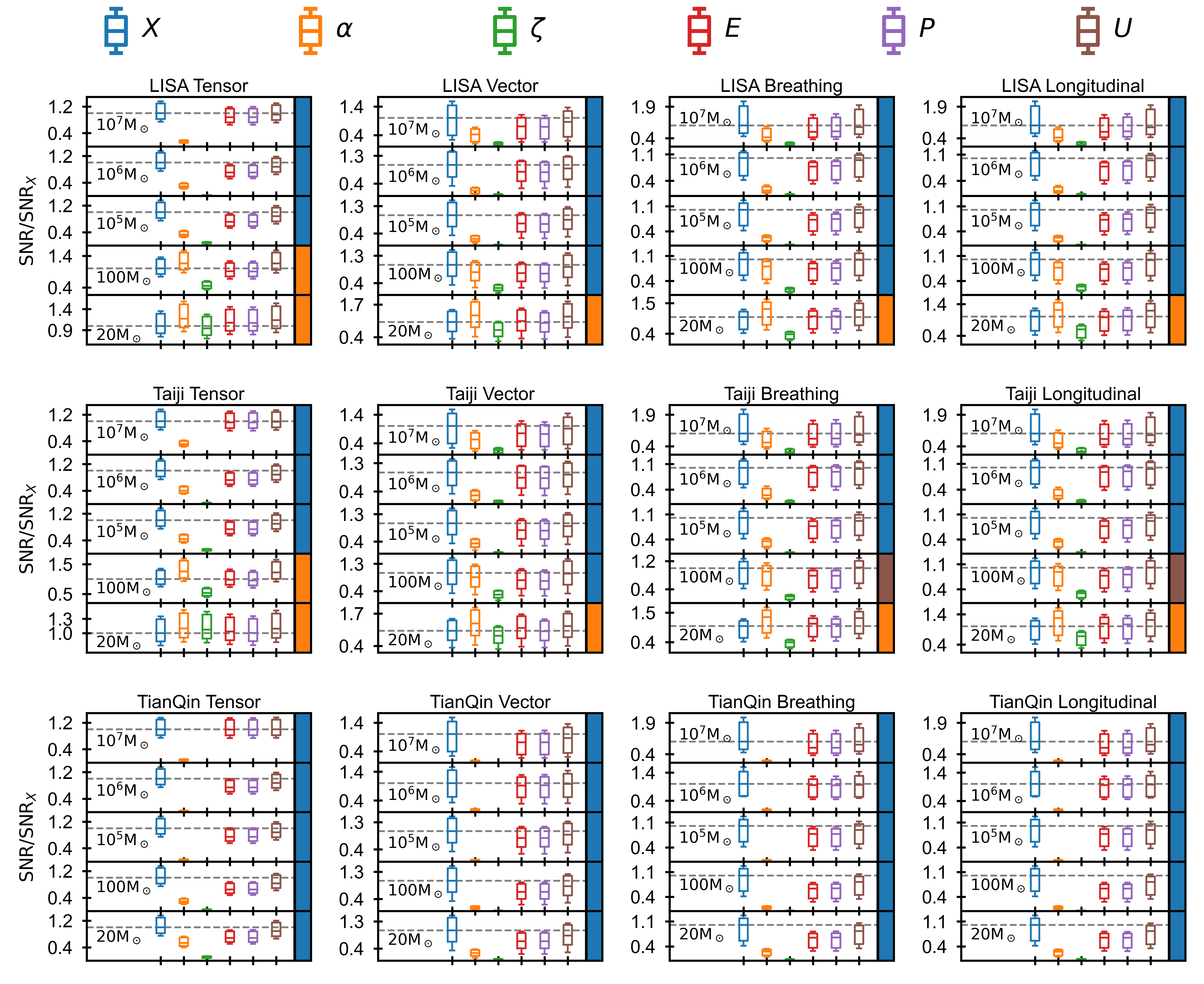}
        \caption{Comparison of SNR for various polarizations under different TDI combinations. This is the result of using three space-based detectors to detect five different mass BBHs at different sky locations. The upper and lower horizontal lines in each box represent the 90\% confidence interval. The edges of the box correspond to the upper and lower quartiles, while the lines inside the box represent the median. The median SNR of the $X$ channel under the same conditions is taken as the benchmark. The color block on the right side of the small graph represents the best TDI combination in this situation.}\label{fig:SNR}
    \end{minipage}
\end{figure*}

To investigate the influences of different TDI combinations on specific GW signals, we simulate the detection of BBH signals. 
As depicted in Fig.~\ref{fig:GW}(a), we select these five BBHs with different masses and utilize the sensitivity curve to calculate the SNR with different parameters. 
The results are presented in Fig.~\ref{fig:SNR}.

For MBHB, the SNR of the $X$ channel is the highest because its frequency band is lower than $f_*$. 
For SBBH, the results are different. 
Since part of the SBBH frequency band is lower than $f_*$ of TianQin, the result of the $X$ channel is still the best for TianQin. 
Considering LISA and Taiji, the entire frequency band of SBBH is higher than $f_*$, so the result of the $X$ channel is not necessarily the best. 
For SBBH with $M = 20\ \mathrm{M}_{\odot }$, the SNR of the $\alpha$ channel is the highest. 
For SBBH with $M = 100\ \mathrm{M}_{\odot }$, the $U$ channel is a suitable choice, even if it is not always the highest, it is very close.

Under the same parametric conditions, the lower the sensitivity curve is, the higher the SNR will be. 
For the Fisher information matrix, a lower sensitivity curve results in a reduction of the parameter uncertainty~\cite{my_paper4,FIM}. 
The results in Figs.~\ref{fig:sensitivity} and \ref{fig:SNR} can reflect the ability to constrain on GW polarizations, and a higher SNR corresponds to a smaller parameter uncertainty. 
In this work, we therefore focus on comparing the relative performance of different TDI combinations through their sensitivity curves and SNRs. We note that a single SNR threshold is not sufficient to claim the detection of additional polarization modes, since their observability depends on the magnitude of the corresponding non GR contributions in the waveform. A detailed discussion on how such constraints are interpreted within the Fisher matrix framework can be found in our previous work~\cite{my_paper4}. % 改

Consequently, in terms of the selected BBHs, when the GW frequency band is lower than $f_*$, the $X$ channel is the best choice; when the GW frequency band is higher than $f_*$, both the $\alpha$ and $U$ channels are excellent options.
In the case of single-arm damage and missing data, the $U$ channel is the optimal selection.

% ---------------------------------------
\subsection{The $\mathcal{AET}$ channels}\label{subsec:AET_channels}
In the previous section, we compare the influences of different TDI combinations on the sensitivity curve and the SNR of BBH. 
It could be observed that the $X$ channel is important. 
Hence, we conduct research on the $\mathcal{AET}$ channels, which are formed by the linear combination of the $XYZ$ channels.

\begin{figure}[ht]
    \begin{minipage}{\columnwidth}
        \centering
        \includegraphics[width=0.9\textwidth,
        trim=0 0 0 0,clip]{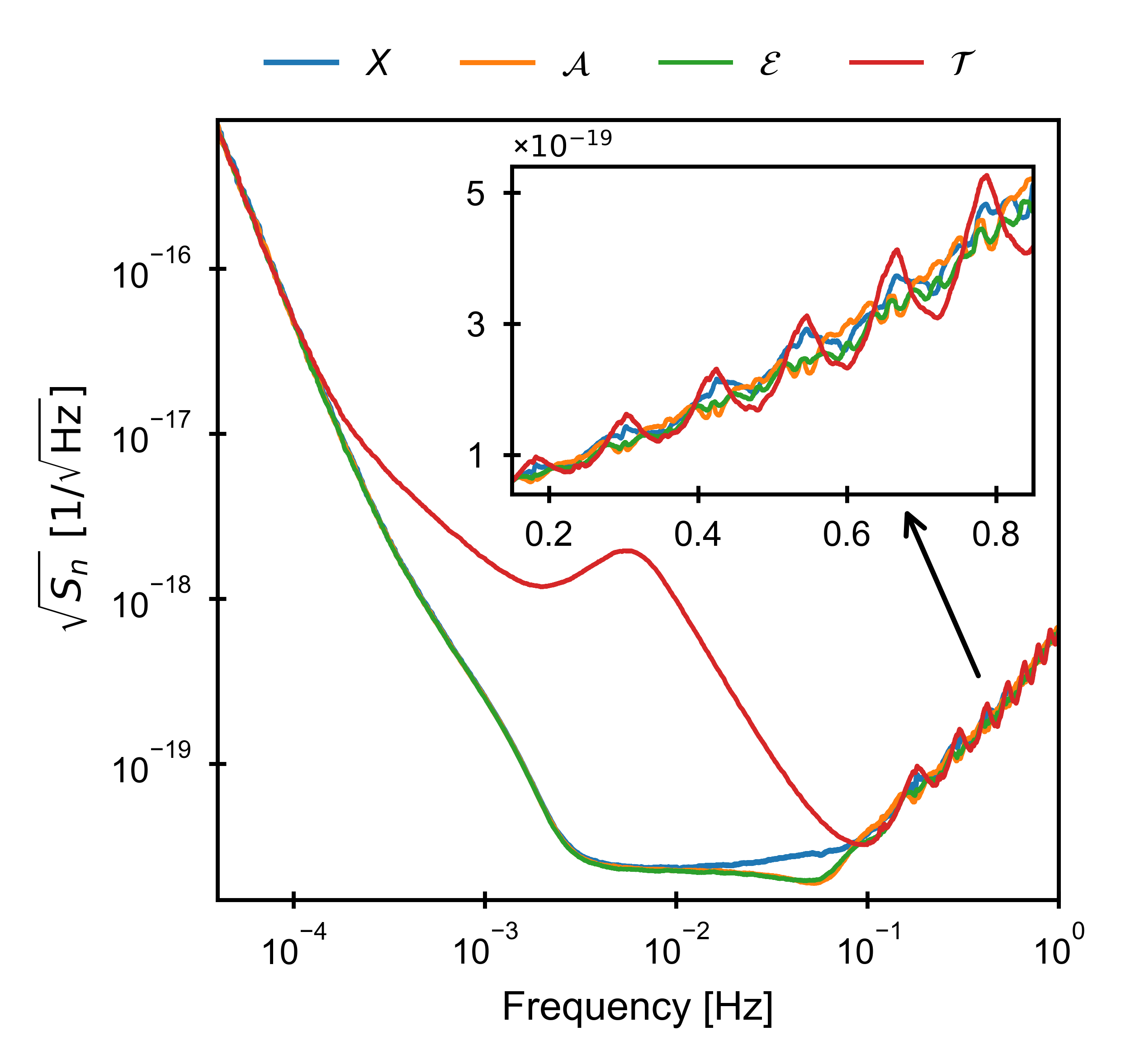}
        \caption{Sensitivity curves for the tensor mode of LISA. }\label{fig:sensitivity_OP}
    \end{minipage}
\end{figure}

In Fig.~\ref{fig:sensitivity_OP}, the sensitivity curves of the $X$, $\mathcal{A}$, $\mathcal{E}$, and $\mathcal{T}$ channels in the LISA tensor mode are depicted. 
As the null channel, the $\mathcal{T}$ channel demonstrates extremely poor sensitivity within the $10^{-4}\sim 10^{-1}$ Hz frequency band, rendering it almost ineffective for signal detection and thereby making it a favorable monitor for certain noises. 
In the frequency band greater than 0.1 Hz, the $\mathcal{T}$ channel undergoes more significant fluctuations compared with the other channels.
The $\mathcal{A}$ and $\mathcal{E}$ channels are highly similar and slightly outperform the $X$ channel in the most sensitive frequency band, particularly in the $10^{-2}\sim 10^{-1}$ Hz frequency band.

\begin{figure}[ht]
    \begin{minipage}{\columnwidth}
        \centering
        \includegraphics[width=0.95\textwidth,
        trim=0 0 0 0,clip]{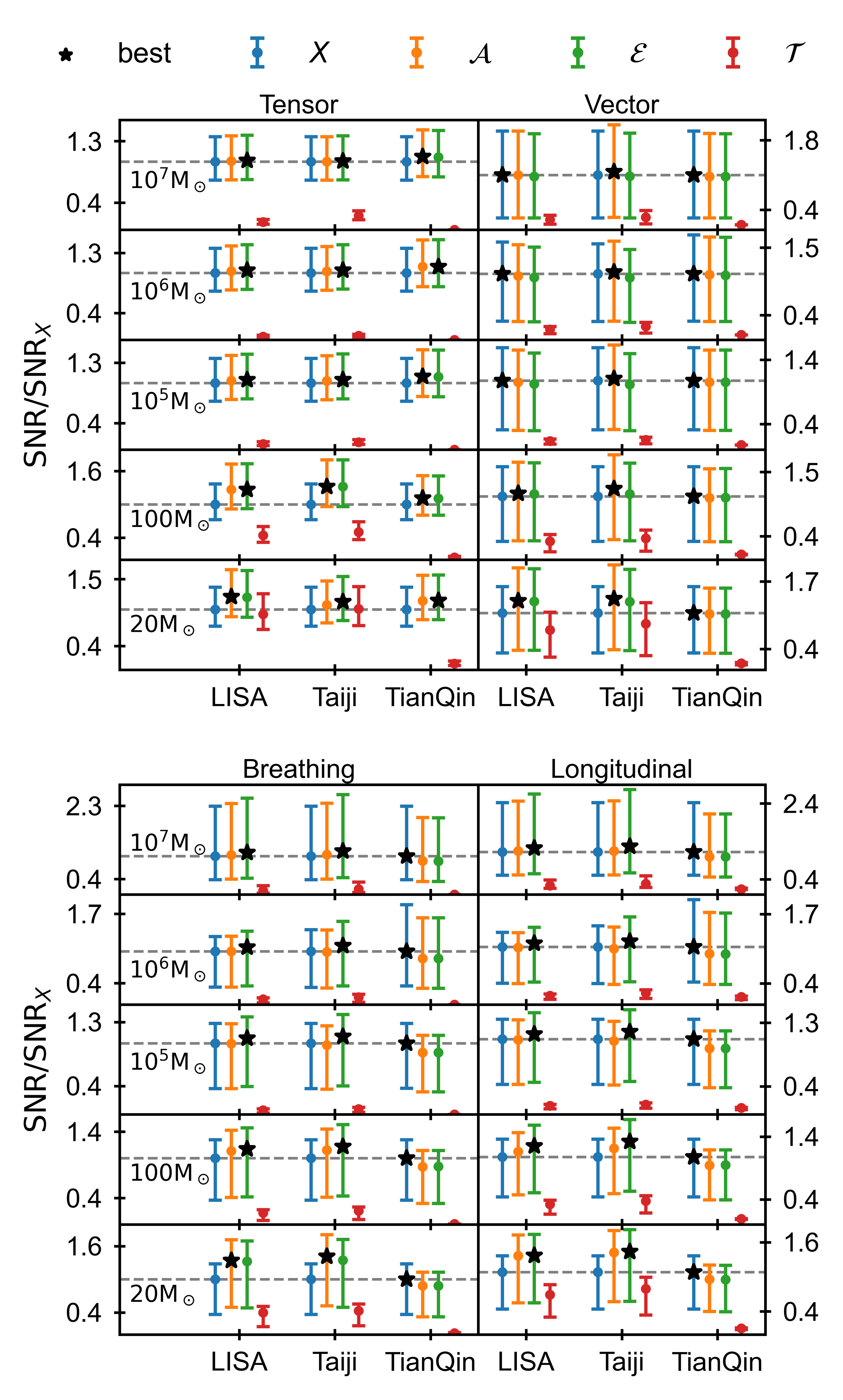}
        \caption{Comparison of SNR for various polarizations under Michelson and optimal combinations. For simplicity, we use points to represent the median, with error bars above and below indicating the 90\% confidence interval. The black stars represent the best channels.}\label{fig:SNR_OP}
    \end{minipage}
\end{figure}

We employ the research approach from the previous section to assess the performance by computing the SNR of BBHs with five different masses. 
The SNR results of the three detectors with different polarizations are presented in Fig.~\ref{fig:SNR_OP}.
For LISA and Taiji, apart from the $\mathcal{T}$ channel, the differences between the $X$, $\mathcal{A}$, and $\mathcal{E}$ channels are negligible. 
Since the sensitivity curves of these channels are highly proximate at low frequencies, the SNR results of MBHB are almost the same. 
In most cases, the results of the $\mathcal{A}$ and $\mathcal{E}$ channels are slightly better than those of the $X$ channel. 
Regarding SBBH, due to the difference in the most sensitive frequency band, the SNR of the $\mathcal{A}$ and $\mathcal{E}$ channels exhibits a marked improvement compared with that of the $X$ channel. 
Even in the tensor mode, for the SBBH with $M = 20\ \mathrm{M}_{\odot }$, the SNR of the $\mathcal{T}$ channel can attain the same level as that of the $X$ channel.

For TianQin, the difference in the sensitive frequency band range leads to different results compared with LISA and Taiji. 
In the tensor mode, the results of the $\mathcal{A}$ and $\mathcal{E}$ channels are slightly better than those of the $X$ channel. 
Among the remaining polarizations, judging from the final results, the $X$ channel turns out to be the best. 
Nevertheless, in the vector mode, the results of the $X$, $\mathcal{A}$, and $\mathcal{E}$ channels are almost the same, while in the breathing and longitudinal modes, the SNR of the $X$ channel is conspicuously higher than that of the $\mathcal{A}$ and $\mathcal{E}$ channels. 
Additionally, in all polarizations, the $\mathcal{T}$ channel is almost unusable.

\begin{figure*}[ht]
    \begin{minipage}{\textwidth}
        \centering
        \includegraphics[width=0.79\textwidth,
        trim=0 0 0 0,clip]{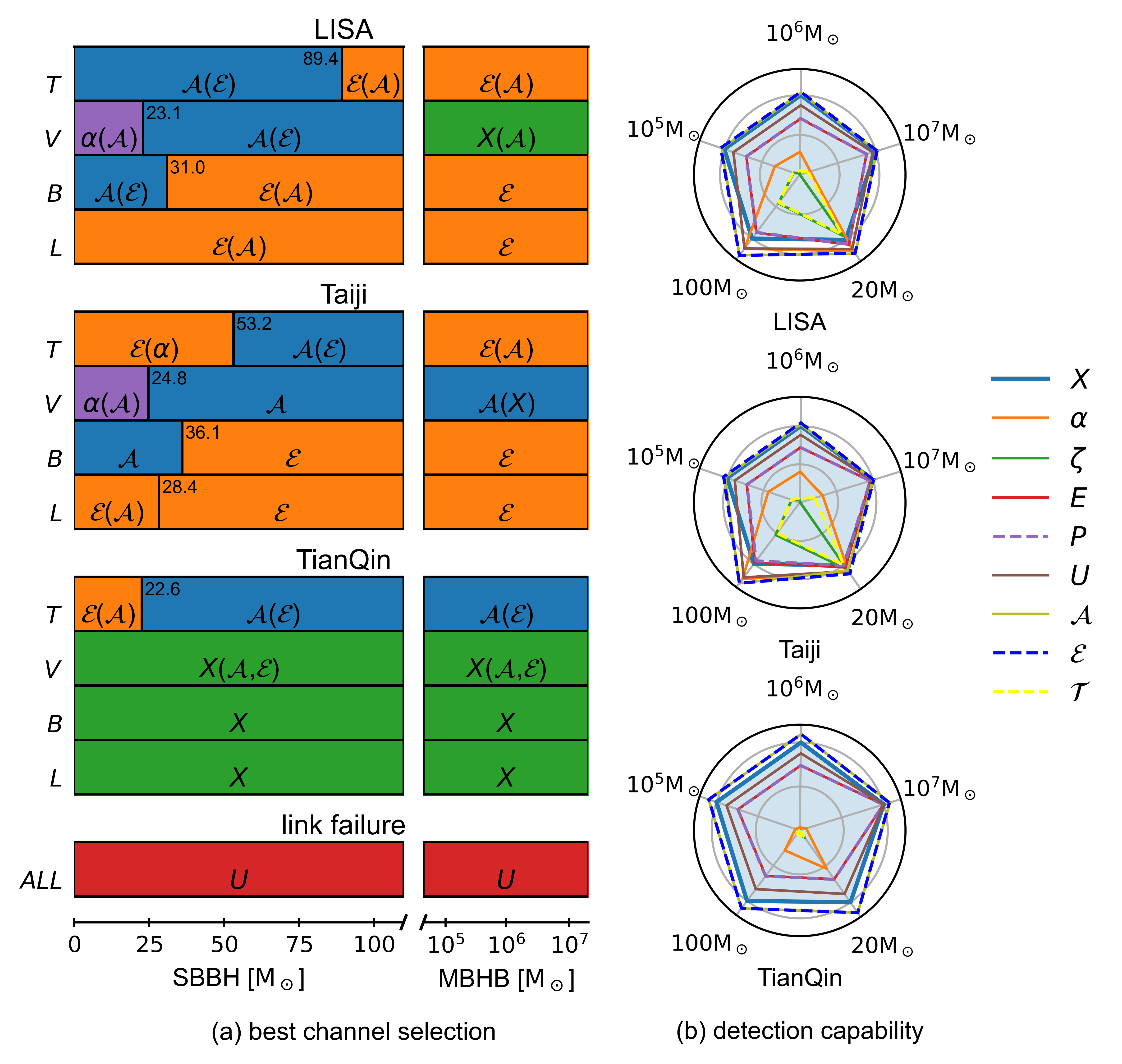}
        \caption{Summary of the best TDI channel selection and detection performance for different GW polarizations across a wide range of BBH masses. (a) shows the best channel selection for detecting different polarizations with SBBH and MBHB. The content before the parentheses represents the best channel in the corresponding situation, and the content within the parentheses indicates the excellent channels, with a deviation of within 3\% from the best channel. The numbers signify the demarcation lines for choosing different channels in the SBBH detection. The left-hand side \textit{T, V, B, L} stands for tensor, vector, breathing, and longitudinal modes, while \textit{ALL} stands for all polarizations. (b) shows the comparison of detection capability for various polarizations under different TDI combinations. The detection capability of the $X$ channel is taken as the benchmark. The further out the line, the greater the detection ability.}\label{fig:TDI_detection}
    \end{minipage}
\end{figure*}

Overall, for different space-based detectors detecting BBHs with different masses, the best TDI combinations are different. 
The specific choices are listed in Fig.~\ref{fig:TDI_detection}(a).
Regarding MBHB, for different detectors to detect different polarizations, there is a sole optimal choice. 
As for SBBH, the best channel needs to be selected based on the mass. 
Under most circumstances, the $X$ channel and the $\mathcal{A}$, $\mathcal{E}$ channels constructed from it have the most powerful performance. 
For LISA and Taiji detecting the vector mode of small-mass SBBH, the result of the $\alpha$ channel is superior to the other channels. 
When one laser link is damaged, the $U$ channel is always the best choice for detecting BBH with any polarization and any mass.

% ---------------------------------------
\subsection{Detection capability}
Considering the influence brought by different TDI combinations, in addition to the study of different polarizations, the detection capability is also of great significance. 
In Secs.~\ref{subsec:Constraints_on_polarization} and \ref{subsec:AET_channels}, the SNRs corresponding to different polarizations are computed. 
These SNRs are expressed as relative values, meaning that the value of the ppE parameter does not influence the results. 
In this section, we consider GW signals that encompass all polarizations, as calculated using Eq.~(\ref{eq:SNR_all}).
For ppE parameters, we set $\alpha_Q=96/5$ and $\alpha_{D,V,B,L}=10^{-3}$~\cite{my_paper4}.
The detection capability is measured by calculating the total SNR under different parameters. 
The results of five BBHs with different masses under various parameters and using different TDI combinations are presented as shown in Fig.~\ref{fig:TDI_detection}(b).

For LISA and Taiji, their orbital configurations are similar, thereby yielding similar results. 
In the detection of MBHB, the results of the $\mathcal{A}$ and $\mathcal{E}$ channels are identical and marginally superior to that of the $X$ channel, while the other channels are inferior to the $X$ channel. 
In the detection of SBBH, the $\mathcal{A}$ and $\mathcal{E}$ channels have the strongest detection ability, and the $\alpha$ and $U$ channels surpass the $X$ channel. 
For SBBH with even lower mass, the $E$ and $P$ channels even exceed the $X$ channel, and the detection capabilities of the $\zeta$ and $\mathcal{T}$ channels have also increased.

For TianQin, the difference in its frequency band results in similar results in detecting MBHB and SBBH.
The detection capabilities of the $\mathcal{A}$ and $\mathcal{E}$ channels are identical and the strongest in all circumstances. 
The result of the $X$ channel comes next, and the result of the $U$ channel is lower than that of the $X$ channel. 
The $E$ and $P$ channels have the same detection capabilities, as do the $\zeta$ and $\mathcal{T}$ channels. 
Additionally, as the mass of BBH decreases, the results of the $\alpha$, $\zeta$ and $ \mathcal{T}$ channels improve.

In conclusion, in terms of detection capabilities, the $\mathcal{A}$ and $\mathcal{E}$ channels are the best in all respects. 
The $X$ channel remains a very good option in most cases, especially for MBHB. 
When detecting SBBH with LISA and Taiji, choosing the $\alpha$ and $U$ channels is appropriate. 
In the $E$, $P$ and $U$ channels, if there is a failure that leads to data loss, the $U$ channel is the best choice.

% =======================================

\subsection{Results for SGWB}
\begin{figure*}[ht]
    \begin{minipage}{\textwidth}
        \centering
        \includegraphics[width=0.9\textwidth,
        trim=0 0 0 0,clip]{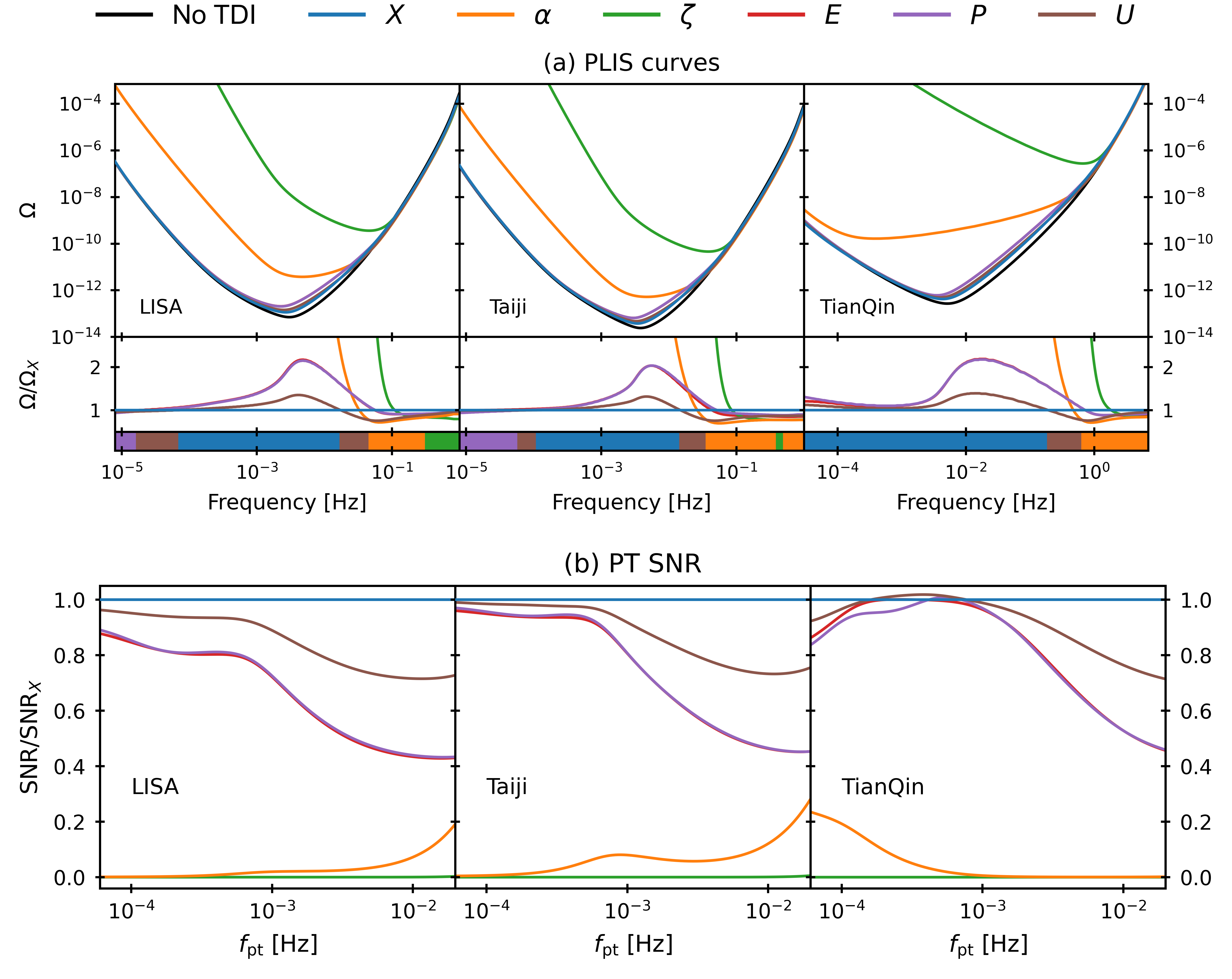}
        \caption{Comparison of the PLIS curves and the SNR for detecting PT spectra across different detectors. (a) shows the PLIS curves for different TDI combinations of each detector, arranged in the same layout as Fig.~\ref{fig:sensitivity}. The first row presents the PLIS curves for all TDI combinations, the second row shows comparisons normalized to the $X$ channel, and the third row highlights the best TDI combinations. (b) displays the SNR for detecting PT spectra, where the SNR of the $X$ channel is taken as the reference. The horizontal axis corresponds to different peak frequencies $f_{\rm pt}$ of the PT spectrum, representing PT models with different underlying parameters.
}\label{fig:PLIS_SNR}
    \end{minipage}
\end{figure*}

In the previous sections, we derive the sensitivity curves for different TDI combinations across multiple detectors and evaluate the SNRs and detection capabilities for BBHs with different masses.
Given the diversity of potential GW sources accessible to future space-based detectors, we extend our analysis beyond astrophysical binaries and investigate the detectability of cosmological sources.
In this subsection, we focus on tensor-polarized SGWB signals and consider two representative classes: power-law backgrounds and first-order PT backgrounds.
Specifically, we construct the PLIS curves for power-law spectra and compute the SNRs for PT spectra with different parameters.
The results are summarized in Fig.~\ref{fig:PLIS_SNR}.
Since our analysis is based on autocorrelation measurements and does not involve overlap reduction functions, we restrict ourselves to TDI combinations with cyclic permutations of indices, and the $\mathcal{AET}$ channels are not considered.

We first discuss the results for power-law SGWB signals.
Figure~\ref{fig:PLIS_SNR}(a) shows the PLIS curves for different TDI combinations.
The trends closely follow those observed in the sensitivity curves shown in Fig.~\ref{fig:sensitivity}.
In the most sensitive frequency band, the $X$ channel consistently provides the best performance.
At higher frequencies, the $\alpha$ and $U$ channels become more competitive.
At lower frequencies, the differences between LISA and Taiji are relatively small, with the $P$ and $U$ channels exhibiting improved sensitivity.
For TianQin, the $X$ channel remains optimal across the middle and low-frequency ranges.

We next turn to SGWB signals generated by first-order PTs.
To facilitate a direct comparison, we compute the SNRs over the parameter range predicted by current theoretical models, with peak frequencies spanning
$f_{\rm pt}=10^{-4}-10^{-2}\,\mathrm{Hz}$.
The corresponding results are shown in Fig.~\ref{fig:PLIS_SNR}(b).
For both LISA and Taiji, the $X$ channel consistently yields the highest SNR over the entire frequency range considered.
For TianQin, the $X$ channel performs best at frequencies above $10^{-3}\,\mathrm{Hz}$, while at lower frequencies the $U$ channel provides a higher SNR.

In conclusion, we have systematically investigated the performance of different TDI combinations for space-based GW detectors in probing both astrophysical and cosmological sources.
By consistently modeling the detector noise, constructing the TDI responses, and evaluating the corresponding sensitivity and PLIS curves, we have quantified the detection capabilities for BBHs and SGWBs under different polarizations.
Our results show that the $X$ channel generally provides the optimal performance across a wide range of scenarios, while other channels, such as $U$ and $\alpha$, can become competitive in specific frequency bands or for particular source classes.
These findings highlight the importance of channel-dependent analyses in space-based GW observations and provide a unified framework for comparing different detectors and TDI configurations.

% =======================================
\section{Conclusion}\label{sec:Conclusion}
In this paper, we investigate the constraints and detection capabilities of GW polarization with space-based GW detectors in different TDI combinations.
Specifically, for space-based detectors, we take into account the orbital configurations of LISA, Taiji, and TianQin. 
Considering the possible circumstances that might occur in reality, seven different second-generation TDI combinations are simulated. 
We conduct simulations of both GW signals and instrumental noise separately. 
The average response is calculated through the flat-spectrum SGWB to obtain the sensitivity curves. 
Utilizing GW waveforms encompassing six polarizations in the ppE framework and considering five different BBH masses, the SNRs under various parameter conditions are evaluated.
In addition, we compute the PLIS curves applicable to power-law SGWB signals and evaluate the SNRs of PT-induced SGWBs over a range of model parameters. 
By employing these methods, we systematically analyze the impact of different TDI combinations on the constraints and detection capabilities of GW polarizations. 

Our research indicates that for the fundamental six TDI combinations, the $X$ channel exhibits superior sensitivity when the frequency is less than $f_*$.
When the frequency exceeds $f_*$, a flexible choice is necessary. 
In the case of single-arm data loss, the $U$ channel demonstrates better sensitivity. 
For MBHB, the SNR of the $X$ channel is the highest. 
For SBBH, the use of the $\alpha$ or $U$ channel is more favorable in LISA and Taiji, while TianQin still performs best with the $X$ channel. 
Considering the $\mathcal{AET}$ channels constructed from the linear combination of the $XYZ$ channels, both the $\mathcal{A}$ and $\mathcal{E}$ channels are excellent options in LISA and Taiji. 
For TianQin, the $\mathcal{A}$ and $\mathcal{E}$ channels are only optimal for the tensor mode, while for additional polarizations, the $X$ channel is still superior.
In contrast to LISA and Taiji, where both the $\mathcal{A}$ and $\mathcal{E}$ channels provide strong capabilities for GW polarization tests, TianQin exhibits a markedly different behavior, with the $X$ channel playing the dominant role in probing GW polarizations. 
This qualitative difference between geocentric and heliocentric space-based detector configurations represents a key result of the present work and has not been explicitly emphasized in previous studies. 

Regarding the detection ability, $\mathcal{A}$ and $\mathcal{E}$ channels are the best, outperforming other TDI combinations. 
In the remaining TDI combinations, the $X$ channel is excellent for detecting MBHB. 
For detecting SBBH with LISA and Taiji, the detection capabilities of $\alpha$ and $U$ channels are better, yet the $X$ channel is still the best with TianQin. 
When detecting SBBH of smaller masses, $E$ and $P$ channels with LISA and Taiji even surpass the $X$ channel, and the detection capabilities of $\zeta$ and $\mathcal{T}$ channels are also close to that of the $X$ channel.
Moreover, if one link is damaged, the $U$ channel has the best detection ability.
Overall, this work consolidates and clearly highlights the essential features of GW polarization sensitivity related to different detectors and TDI channels that were only implicitly contained in previous studies.

For SGWB signals, the PLIS analysis for power-law backgrounds shows that the relative performance of different TDI combinations closely follows the trends inferred from the sensitivity curves, with the $X$ channel offering the best overall performance in most frequency ranges, while the $U$ channel becomes competitive in specific bands.
For SGWBs generated by first-order PTs, the $X$ channel generally yields the highest SNR for LISA and Taiji across the parameter space considered.
For TianQin, the $X$ channel is optimal at higher peak frequencies, whereas the $U$ channel provides improved sensitivity at lower peak frequencies.
These results demonstrate that the optimal TDI choice for SGWB detection depends on both the spectral shape of the background and the frequency band.

In future research, we plan to explore a range of TDI combinations to assess their impact on various GW signals, including different SGWB signals. 
By incorporating these diverse TDI techniques, we aim to enhance the thoroughness of our study and achieve more realistic, robust results.
This expanded approach will not only provide a more comprehensive analysis of the detection capabilities of space-based GW detectors but also improve our ability to search for alternative polarization modes, offering a deeper test of GR. 
Ultimately, our research will contribute valuable insights into the performance of different TDI combinations and advance the detection of GW across different detectors.

\begin{acknowledgements}
This work was supported by the National Key Research and Development Program of China (Grant No. 2023YFC2206702), the National Natural Science Foundation of China (Grant Nos. 12575072, 12547101 and 125B2102), the Fundamental Research Funds for the Central Universities Project (Grant No. 2024IAIS-ZD009), and the Natural Science Foundation of Chongqing (Grant No. CSTB2023NSCQ-MSX0103).
\end{acknowledgements}

\appendix
\section{TDI combinations}\label{sec:TDI_conbinations}
In the appendix, we give all the second-generation TDI combination definitions used in this study.
For details of the derivation of these specific forms, see Refs.~\cite{PhD,TDI_combination1,TDI_combination2}.

To more succinctly and effectively represent the TDI combination, we introduce the time-advance operator $D_{\bar{i}}$, which is defined as the inverse of the time-delay operator $D_{i }$ presented in Eq.~(\ref{eq:D}). Specifically, the time advance operator can be formally defined as~\cite{TDI_all}
\begin{equation}
    D_{\bar{i}}\eta(t)=\eta(t+L_i(t)).
\end{equation}

\textit{No TDI} refers to the interference only between adjacent links:
\begin{equation}
    \mathit{No\ TDI}=\eta_1-\eta_{1'} ,
\end{equation}
where the specification for the use of indicators is detailed in Appendix D of Ref.~\cite{PhD}.
All the second-generation TDI combinations used are as follows:

\begin{enumerate}
\item the Michelson combination ($X,Y,Z$):
\begin{equation}
    \begin{aligned}
        X & =\eta_1+D_3\eta_{2'}+D_{33'}\eta_{1'}+D_{33'2'}\eta_3+D_{33'2'2}\eta_{1'} \\
         & +D_{33'2'22'}\eta_3+D_{33'2'22'2}\eta_1+D_{33'2'22'23}\eta_{2'} \\
         & -(\eta_{1'}+D_{2'}\eta_{3}+D_{2'2}\eta_{1}+D_{2'23}\eta_{2'}+D_{2'233'}\eta_{1} \\
         & +D_{2'233'3}\eta_{2'}+D_{2'233'33'}\eta_{1'}+D_{2'233'33'2'}\eta_{3}).
    \end{aligned}
\end{equation}

\item the Sagnac combination ($\alpha, \beta, \gamma $):
\begin{equation}
    \begin{aligned}
        \alpha & =(D_{2'1'3'}-1)(\eta_{1}+D_{3}\eta_{2}+D_{31}\eta_{3}) \\
        & -(D_{312}-1)(\eta_{1'}+D_{2'}\eta_{3'}+D_{2'1'}\eta_{2'}).
    \end{aligned}
\end{equation}

\item the fully symmetric combination ($\zeta $):
\begin{equation}
    \begin{aligned}
        \zeta & =(D_{2'3'}-D_{1})(D_{3}\eta_{3}-D_{3}\eta_{3'}+D_{1'}\eta_{1}) \\
        & -(D_{32}-D_{1'})(D_{1}\eta_{1'}-D_{2'}\eta_{2}+D_{2'}\eta_{2'}).
    \end{aligned}
\end{equation}

\item the Beacon combination ($P, Q, R$):
\begin{equation}
    \begin{aligned}
        P & =(1-D_{2'\bar{1}3'31\bar{2}'}+D_{33'2'\bar{1}3'}-D_{2'\bar{1}3'})\eta_{1} \\
        & +(-1+D_{33'}+D_{2'\bar{1}3'31\bar{2}'}-D_{33'2'\bar{1}3'31\bar{2}'})\eta_{1'} \\
        & +(D_{33'2'\bar{1}3'3}-D_{33'2'\bar{1}}-D_{2'\bar{1}3'3}+D_{2'\bar{1}})\eta_{2} \\
        & +(D_3+D_{33'2'\bar{1}}-D_{2'\bar{1}}-D_{2'\bar{1}3'31\bar{2}'3})\eta_{2'}.
    \end{aligned}
\end{equation}

\item the Monitor combination ($E, F, G$):
\begin{equation}
    \begin{aligned}
        E & =(1-D_{2^{\prime}1^{\prime}1\bar{2}^{\prime}})[(1-D_{311^{\prime}\bar{3}})\eta_{1}+D_{3}\eta_{2}+D_{31}\eta_{3^{\prime}}] \\
        & -(1-D_{311^{\prime}\bar{3}})[(1-D_{2^{\prime}1^{\prime}1\bar{2}^{\prime}})\eta_{1^{\prime}}+D_{2^{\prime}}\eta_{3^{\prime}}+D_{2^{\prime}1^{\prime}}\eta_{2}].
    \end{aligned}
\end{equation}

\item the Relay combination ($U, V, W$):
\begin{equation}
    \begin{aligned}
        U& =(D_{33^{\prime}2^{\prime}1^{\prime}3}-1)(\eta_{1^{\prime}}+D_{2^{\prime}}\eta_{3^{\prime}}+D_{2^{\prime}1^{\prime}}\eta_{2^{\prime}}+D_{2^{\prime}1^{\prime}3^{\prime}}\eta_{1}) \\
        & -(D_{2^{\prime}1^{\prime}3^{\prime}}-1)D_{3}(\eta_{2^{\prime}}+D_{3^{\prime}}\eta_{1^{\prime}}+D_{3^{\prime}2^{\prime}}\eta_{3^{\prime}}) \\
        & -(D_{33^{\prime}2^{\prime}1^{\prime}\bar{3}}-1)\eta_{1}.
    \end{aligned}
\end{equation}

\end{enumerate}

The $\mathcal{A},\mathcal{E}$, and $\mathcal{T}$ channels are derived through a linear combination of the $X,Y$, and $Z$ channels (see Eq.~(\ref{eq:AET})).
The other channels can be generated by cyclic permutation of indices: $1\to2\to3\to1$.

\bibliography{references}
\end{document}